\documentclass[fleqn,usenatbib,useAMS]{mnras}

\usepackage{graphicx}	
\usepackage{amsmath}	
\usepackage{amssymb}	
\usepackage{multicol}        
\usepackage{bm}		
\usepackage{pdflscape}	
\usepackage{siunitx}
\usepackage{xcolor}
\usepackage{pdflscape}
\usepackage{multirow}
\usepackage{caption}
\usepackage{datetime2}
\usepackage{scalerel} 
\usepackage{mhchem}
\usepackage{cancel}

\newcommand{\comment}[1]{}  

\DeclareCaptionLabelFormat{continued}{Figure#1~#2 (continued)}
\usepackage[T1]{fontenc}
\usepackage{ae,aecompl}

\usepackage{newtxtext,newtxmath}

\title[SMARTY, a NIR stellar library]{SMARTY: The mileS Moderate resolution neAr-infRared sTellar librarY}

\author[Michele]{
Michele Bertoldo-Coêlho$^{1}$\thanks{E-mail: michele.bertoldo@ufrgs.br},
Rogério Riffel$^{1,2}$,             
Marina Trevisan$^{1}$,              
Natacha Zanon Dametto$^{1,2}$,      
Luis \newauthor Dahmer-Hahn$^{4}$,             
Paula Coelho$^{5}$,                 
Lucimara Martins$^{6}$,             
Daniel Ruschel-Dutra$^{7}$,         
Alexandre Vazdekis$^{2,3}$,         
\newauthor
Alberto Rodríguez-Ardila$^{8,9}$,  
Ana L. Chies-Santos$^{1}$,             
Rogemar A. Riffel$^{10}$,            
Francesco  La Barbera$^{11}$,          
\newauthor
Ignacio Martín Navarro$^{2,3}$,     
Jesus Falcon Barroso$^{2,3}$,        
Tatiana Moura$^{5}$                
\\
$^{1}$Departamento de Astronomia, Instituto de Física, Universidade Federal do Rio Grande do Sul, Av. Bento Gonçalves 9500, 91501-970 Porto Alegre, RS, Brazil \\
$^{2}$Instituto de Astrofísica de Canarias, E-38205 La Laguna, Tenerife, Spain \\
$^{3}$Universidad de La Laguna, Dpto. Astrofísica, E-38206 La Laguna, Tenerife, Spain \\
$^{4}$Shanghai Astronomical Observatory, Chinese Academy of Sciences, 80 Nandan Road, Shanghai 200030, China \\
$^{5}$Universidade de São Paulo, Instituto de Astronomia, Geofísica e Ciências Atmosféricas, Rua do Matão 1226, 05508-090, São Paulo, Brazil \\
$^{6}$NAT – Universidade Cidade de São Paulo/Universidade Cruzeiro do Sul, Rua Galvão Bueno 868, São Paulo-SP, 01506-000, Brazil \\
$^{7}$Departamento de Física, Universidade Federal de Santa Catarina, P.O. Box 476, 88040-900, Florianópolis, SC, Brazil \\
$^{8}$Laboratório Nacional de Astrofísica. Rua dos Estados Unidos, 154, 37504-364 Itajubá, MG, Brazil \\
$^{9}$Instituto Nacional de Pesquisas Espaciais, Av. dos Astronautas, 1758 - Jardim da Granja São José dos Campos/SP - 12227-010, Brazil \\
$^{10}$Departamento de Física, Centro de Ciências Naturais e Exatas, Universidade Federal de Santa Maria, 97105-900, Santa Maria, RS, Brazil \\
$^{11}$INAF - Osservatorio Astronomico di Capodimonte, sal. Moiariello 16, Napoli, 80131, Italy \\
}

\date{}

\pubyear{2024}

\begin{document}
\label{firstpage}
\pagerange{\pageref{firstpage}--\pageref{lastpage}}
\maketitle

\begin{abstract}
Most of the observed galaxies cannot be resolved into individual stars and are studied through their integrated spectrum using simple stellar populations (SSPs) models, with stellar libraries being a key ingredient in building them. Spectroscopic observations are increasingly being directed towards the near-infrared (NIR), where much is yet to be explored. SSPs in the NIR are still limited, and there are inconsistencies between different sets of models. One of the ways to minimize this problem is to have reliable NIR stellar libraries. The main goal of this work is to present {\sc smarty} (mileS Moderate resolution neAr-infRared sTellar librarY) a $\sim0.9-2.4$~\micron\, stellar spectral library composed of 31 stars observed with the Gemini Near-IR Spectrograph (GNIRS) at the $8.1$~m Gemini North telescope and make it available to the community. The stars were chosen from the {\sc miles} library, for which the atmospheric parameters are reliable (and well tested), to populate different regions of the Hertzsprung–Russell (HR) diagram.
Furthermore, five of these stars have NIR spectra available that we use to assess the quality of {\sc smarty}. The remaining 26 stars are presented for the first time in the NIR. We compared the observed {\sc smarty} spectra with synthetic and interpolated spectra, finding a mean difference of $\sim 20\%$ in the equivalent widths and $\sim$1\% in the overall continuum shape in both sets of comparisons. We computed the spectrophotometric broadband magnitudes and colours and compared them with the 2MASS ones, resulting in mean differences up to $0.07$ and $0.10$\,mag in magnitudes and colours, respectively. In general, a small difference was noted between the \textsc{smarty} spectra corrected using the continuum from the interpolated and the theoretical stars. 

\end{abstract}

\begin{keywords}
stars: general -- galaxies: stellar content -- catalogs -- techniques: spectroscopic
\end{keywords}



\section{Introduction}

Galaxies in the local universe are the final product of a very long process, which depends on a combination of internal (e.g. star formation, stellar and chemical evolution) and external processes (e.g. environment). They display a wide range of properties, such as luminosities, stellar masses, gas, and dust content \citep[e.g.][]{Conselice+14, Sanchez+18,Sanchez+20,Sanchez+21,Riffel+21c,Riffel+22,Riffel+23}. The determination of many of these properties relies on the correct characterisation of their stellar content, which, in turn, depends on reliable stellar population models. One of the most common ways of modelling integrated stellar populations is through spectral fitting, which can, for example, combine simple stellar populations (SSPs) in different proportions to build the composite stellar population that best describes the observed galaxy \citep[][]{Tinsley+68,Cid+05,Walcher+11,Conroy+13,Gomes+17,Cappellari+23}. Thus, the SSPs are the most important ingredient in this type of characterisation.

To build up reliable SSP models, one needs several ingredients. Stellar libraries are one of the most fundamental ones \citep[e.g.][and references therein]{Worthey+92,McWilliam+97,Sanchez+06,Gustafsson+08,Husser+13,Coelho+14,Chen+14,Villaume+17,Knowles+21}, and can be either theoretical or empirical. Empirical libraries depend on existing observed stellar spectra, for which high signal-to-noise data can be obtained only for nearby stars. Thus, empirical libraries are restricted to nearby and bright objects, leading to libraries biased to metallicity and abundance ratios of stars in the solar neighbourhood. Besides, these libraries have limited coverage of atmospheric parameters and spectral resolution. Theoretical libraries, on the other hand, can encompass a wide range of parameters, including the possibility of high-resolution spectra. However, these libraries depend on our knowledge of the physics of stellar atmospheres and data of atomic and molecular opacities. Thus, both are important since they complement each other, not only in their use but in their assembly. For instance, theoretical libraries are tested and calibrated using empirical libraries \citep{Coelho+09,Arentsen+19, Coelho+20,Lancon+21} while empirical libraries rely on theoretical libraries for estimating the atmospheric parameters of stars. Both types of libraries have been intensely evolving in the last years and, at least in the optical, can confidently be used to reproduce the integrated spectra of stellar systems (\citealp{Martins+19}; see also \citealp{Moura+19}; \citealp{Renno+20}).

Ideally, for a given set of atmospheric parameters, theoretical and empirical spectra should be equivalent; however, a complete understanding of stellar physics (as well as atmospheric, atomic, and molecular parameters) and greater observational power would be necessary for this scenario to be achieved. A way to overcome this is to use combined empirical and theoretical stellar libraries. For instance, \citet{Westera+02} used empirical data to correct the spectral energy distributions of the BaseL Stellar Library (BaSeL). \citet{Coelho+14} used observed stellar spectra to test her theoretical stellar library. Empirical and theoretical stellar libraries can also be used to build semi-empirical stellar population synthesis models in methods called differential stellar population \citep{Walcher+09} and flexible SPS \citep[FSPS,][]{Conroy+09}, as well as in abundance ratio variations \citep[e.g.][]{Knowles+21,Knowles+23}.

Studies of the stellar content of galaxies using the near-infrared (NIR) bands can date back to the 80's \citep{Rieke+80}. However, they are becoming more popular over the last years \citep[e.g.][]{Origlia+97,Maraston+05,Riffel+06,Riffel+07,Silva+08,Riffel+08,Riffel+09,Riffel+11a,Riffel+11b,Maraston+11,Kotilainen+12,Vazdekis+12,Zibetti+13,Martins+13a,Martins+13b,Dametto+14,Riffel+15,Vazdekis+16,Rock+16,Rock+17,DahmerHahn+18,DahmerHahn+19,Riffel+19,Eftekhari+21,Eftekhari+22b,Riffel+22,Gasparri+21,Gasparri+24} mostly because the detectors have improved in this spectral region and telluric corrections became more efficient \citep[e.g.][]{Smette+15}.  

Additionally, understanding the stellar populations using the NIR spectral region is now of utmost importance since the JWST is producing amazing data in this spectral region \citep[e.g.][]{Luhman+24,Marino+24,Boyett+24}. This spectral region is interesting since it is less affected by dust extinction than the optical, allowing the study of the light content inside optically obscured regions \citep[e.g.][and references therein]{Riffel+06,Riffel+15,Riffel+19}. Besides that, models have predicted that cold evolved stars dominate the NIR emission in galaxies. In particular, the thermally pulsing asymptotic giant branch (TP-AGB), phase of cold, intermediate-mass giants stars of difficult modelling, is believed to contribute up to 80\% in $K$ band luminosity for intermediate-age populations ($0.2-2$\,Gyr) \citep{Maraston+06,Salaris+14}, but with a limited impact on the spectral features \citep{Riffel+15,Rock+17,Eftekhari+21,Eftekhari+22b}. See, for example,  \citet{Verro+22+SSP} for a discussion on the effect of these stars in the building of SSP models.

The path towards developing robust SSP models involves comparing empirical and synthetic stellar spectral libraries across the wavelength ranges of photospheric emission. For instance, the theory of stellar physics enters all SSP models, even when this is only implicit in the association of fundamental stellar parameters with empirical spectral library stars \citep[][]{Lancon+21}. To shed some light on our understanding of the stellar populations in the NIR spectral regions, two aspects are thus fundamental: $i$) expand the existing stellar libraries on the NIR since they do not completely populate the HR diagram \citep[e.g.][]{Cushing+05,Rayner+09,MenesesGoytia+15,Rock+16,Villaume+17,Lancon+21} and $ii$) to fine-tune the NIR theoretical libraries, comparing them with the empirical stellar libraries \citep[e.g.][]{Coelho+14}.

Aimed at helping to overcome these problems, here we present a new set of NIR stellar spectra of a sub-sample of {\sc miles}  stars \citep{Sanchez+06,Falcao+11}, whose atmospheric parameters have been previously determined \citep{Cenarro+07,Garcia-Perez+21}. For this purpose, we selected a sub-sample of stars from the {\sc miles}  stellar library, which have been used to test theoretical stellar spectra in the optical region by \citet[][see \S~\ref{sec:data} for more details]{Coelho+14}. For the selected {\sc miles}  stars, we obtained the NIR data using the Gemini Near-IR Spectrograph (GNIRS) on the Gemini North telescope, from $\sim 0.9$ to $\sim 2.4$~\micron\, at a moderate spectral resolution ($R\sim 1\,300$).  

This paper is organised as follows: in Sect.~\ref{sec:data}, we describe the sample selection, our Gemini observations and data reduction. In Sect.~\ref{sec:results}, we describe the processes applied to calibrate the flux in order to fine-tune the spectra quality. Finally, our last remarks are given in Sect.~\ref{sec:conclusions}.

\section{Data} \label{sec:data}

\begin{figure}
    \centering
    \includegraphics[width=0.48\textwidth]{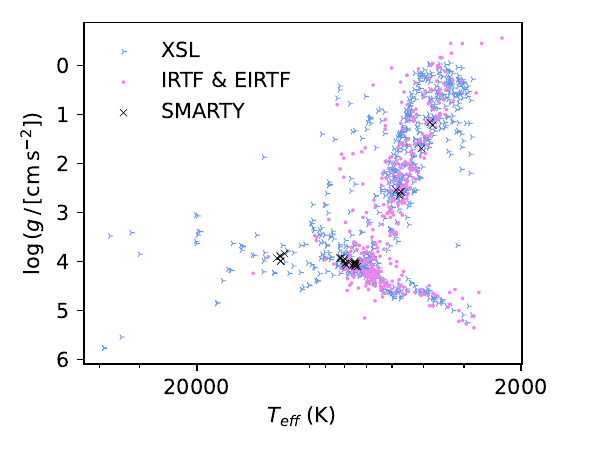}
    \caption{Parameter coverage of existing empirical NIR stellar libraries in the $\log g$ vs. $T_{\rm eff}$ plane. The blue markers and the violet dots are the stars from XSL \citep[683 stars with $R\sim10\,000$;][]{Chen+14,Gonneau+20,Verro+22+XSL} and IRTF+EIRTF \citep[210+287 stars with $R\sim 2\,000$;][]{Cushing+05,Rayner+09,Villaume+17}, respectively. The stars presented in this work are indicated by the black crosses (31 stars with $R\sim 1\,300$).}
    \label{fig:NIRlibraries}
\end{figure}

\subsection{Sample Selection}

Our sample selection was based on a systematic comparison between the {\sc miles} optical observations and a grid of synthetic spectra available in \citet{Coelho+14}. This author grouped the {\sc miles} stars in bins of $T_\textrm{eff}$ and $\log{g}$ with widths given by the uncertainties in these parameters. Our intention was to have from 6 to 8 stars for each $\log{g}$ and [Fe/H] with effective temperature varying around 100\,K to help correct absorption lines of synthetic spectra. However, due to limitations of observing time, we were able to observe only a sub-sample of these stars. We selected the stars giving priority to hotter stars, which are lacking in the IRTF \citep[Infrared Telescope Facility Spectral Library; contains 210 stars within $0.8 - 5.4$\,\micron\ with medium-resolution $R\sim 2\,000$ observed with the cross-dispersed spectrograph SpeX from the NASA Infrared Telescope Facility on Mauna Kea, Hawaii;][]{Cushing+05,Rayner+09} and EIRTF libraries \citep[Extended IRTF Spectral Library; contains 287 stars observed with the SpeX within $0.7 - 2.5$\,\micron\ with $R\sim 2\,000$;][]{Villaume+17}, and also to have a diverse distribution on the HR diagram. 

The final sample comprises 31 stars listed in Tab.~\ref{tab:stars}. The sample is well distributed in the atmospheric parameter space (see Fig.~\ref{fig:NIRlibraries}) and with data available in the optical. Of this sample, 5 stars are common with other NIR libraries \citep[IRTF, EIRTF, and XSL, the X-shooter Spectral Library; contains 683 stars within $0.35 - 2.48$\,\micron\ with moderate-to-high resolution $R\sim10\,000$ observed with the X-shooter three-arm spectrograph of the Very Large Telescope on Cerro Paranal, Chile;][]{Chen+14,Gonneau+20,Verro+22+XSL} to be used as a control sample, and 26 are observed for the first time in the NIR. The spectra in the optical region ($3\,525 - 7\,500$\,\AA) are available in the {\sc miles} library \citep{Sanchez+06,Falcao+11} with a resolution of 2.5\,\AA\  (in FWHM), while the NIR spectral range was observed with GNIRS at $R\sim1\,300$. A comprehensive description of the observation process and data reduction is provided in \S~\ref{sub:observation}.

\begin{table}
\renewcommand{\tabcolsep}{1.5mm}
\renewcommand{\arraystretch}{0.9}
    \centering
    \caption{Stellar atmospheric parameters and other information for the \textsc{smarty} stars.}
    \begin{tabular}{crrrcc}
    \hline 
    \textbf{\footnotesize{Star}} &
    \multicolumn{1}{c}{\scalebox{0.9}{${T_\textrm{eff}}$}} & 
    \multicolumn{1}{c}{\scalebox{0.9}{${\log{g}}$}} &
    \multicolumn{1}{c}{\scalebox{0.9}{$\textrm{[Fe/H]}$}} & 
    \multicolumn{1}{c}{\scalebox{0.9}{${\rm E}({\rm B}-{\rm V})$}} &  
    \textbf{\footnotesize{Spec. Type}}  \\
     & 
    \multicolumn{1}{c}{\scalebox{0.76}{$\mathrm{K}$}} & 
    \multicolumn{1}{c}{\scalebox{0.76}{$\log{[\mathrm{cm}\,\mathrm{s}^{-2}]}$}} & 
    \multicolumn{1}{c}{} &  
    \multicolumn{1}{c}{} & 
       \\
    \textbf{\tiny{(1)}} & 
    \multicolumn{1}{c}{\textbf{\tiny{(2)}}} & 
    \multicolumn{1}{c}{\textbf{\tiny{(3)}}} & 
    \multicolumn{1}{c}{\textbf{\tiny{(4)}}} &  
    \multicolumn{1}{c}{\textbf{\tiny{(5)}}} & 
    \textbf{\tiny{(6)}}   \\ \hline

\footnotesize{HD 015798}     &  \footnotesize{6527}   &    \scalebox{0.9}{4.07}  &    \scalebox{0.9}{-0.12}   &  \scalebox{0.9}{0.000}   &  \footnotesize{ F5V                   }   \\
\footnotesize{HD 026322}     &  \footnotesize{7008}   &    \scalebox{0.9}{3.94}  &    \scalebox{0.9}{0.13}    &  \scalebox{0.9}{0.003}   &  \footnotesize{ F2IV-V                }   \\
\footnotesize{HD 027295$^\mathrm{a}$}     &  \footnotesize{11034}  &    \scalebox{0.9}{3.99}  &    \scalebox{0.9}{-0.11}   &  \scalebox{0.9}{0.000}   &  \footnotesize{ B9IV                  }   \\
\footnotesize{HD 029375}     &  \footnotesize{7240}   &    \scalebox{0.9}{3.93}  &    \scalebox{0.9}{0.13}    &  \scalebox{0.9}{0.053}   &  \footnotesize{ F0V                   }   \\
\footnotesize{HD 071030}     &  \footnotesize{6541}   &    \scalebox{0.9}{4.03}  &    \scalebox{0.9}{-0.15}   &  \scalebox{0.9}{0.000}   &  \footnotesize{ F6V                   }   \\
\footnotesize{HD 078234}     &  \footnotesize{6976}   &    \scalebox{0.9}{4.04}  &    \scalebox{0.9}{-0.06}   &  \scalebox{0.9}{0.014}   &  \footnotesize{ F2V                   }   \\
\footnotesize{HD 087822$^\mathrm{b}$}     &  \footnotesize{6573}   &    \scalebox{0.9}{4.06}  &    \scalebox{0.9}{0.10}    &  \scalebox{0.9}{0.018}   &  \footnotesize{ F4V                   }   \\
\footnotesize{HD 113022}     &  \footnotesize{6491}   &    \scalebox{0.9}{4.09}  &    \scalebox{0.9}{0.11}    &  \scalebox{0.9}{0.000}   &  \footnotesize{ F6Vs                  }   \\
\footnotesize{HD 114642}     &  \footnotesize{6491}   &    \scalebox{0.9}{4.04}  &    \scalebox{0.9}{-0.04}   &  \scalebox{0.9}{0.000}   &  \footnotesize{ F6V                   }   \\
\footnotesize{HD 121299}     &  \footnotesize{4695}   &    \scalebox{0.9}{2.58}  &    \scalebox{0.9}{0.10}    &  \scalebox{0.9}{0.019}   &  \footnotesize{ K2III                 }   \\
\footnotesize{HD 137391$^\mathrm{c}$}     &  \footnotesize{7186}   &    \scalebox{0.9}{3.93}  &    \scalebox{0.9}{0.10}    &  \scalebox{0.9}{0.004}   &  \footnotesize{ F0V                  }   \\
\footnotesize{HD 142908}     &  \footnotesize{7038}   &    \scalebox{0.9}{3.98}  &    \scalebox{0.9}{-0.02}   &  \scalebox{0.9}{0.004}   &  \footnotesize{ F0IV                  }   \\
\footnotesize{HD 143807}     &  \footnotesize{10727}  &    \scalebox{0.9}{3.84}  &    \scalebox{0.9}{-0.01}   &  \scalebox{0.9}{0.024}   &  \footnotesize{ A0p...                }   \\
\footnotesize{HD 145976$^\mathrm{c}$}     &  \footnotesize{6927}   &    \scalebox{0.9}{4.08}  &    \scalebox{0.9}{-0.02}   &  \scalebox{0.9}{0.018}   &  \footnotesize{ F3V                   }   \\
\footnotesize{HD 149121}     &  \footnotesize{11099}  &    \scalebox{0.9}{3.89}  &    \scalebox{0.9}{0.03}    &  \scalebox{0.9}{0.007}   &  \footnotesize{ B9.5III               }   \\
\footnotesize{HD 155078}     &  \footnotesize{6508}   &    \scalebox{0.9}{4.00}  &    \scalebox{0.9}{0.03}    &  \scalebox{0.9}{0.032}   &  \footnotesize{ F5IV                  }   \\
\footnotesize{HD 157856}     &  \footnotesize{6523}   &    \scalebox{0.9}{4.04}  &    \scalebox{0.9}{-0.07}   &  \scalebox{0.9}{0.000}   &  \footnotesize{ F3V                   }   \\
\footnotesize{HD 166285}     &  \footnotesize{6389}   &    \scalebox{0.9}{4.10}  &    \scalebox{0.9}{-0.06}   &  \scalebox{0.9}{0.000}   &  \footnotesize{ F5V                   }   \\
\footnotesize{HD 169027}     &  \footnotesize{11030}  &    \scalebox{0.9}{3.89}  &    \scalebox{0.9}{-0.08}   &  \scalebox{0.9}{0.031}   &  \footnotesize{ A0                    }   \\
\footnotesize{HD 172103}     &  \footnotesize{6815}   &    \scalebox{0.9}{4.01}  &    \scalebox{0.9}{0.03}    &  \scalebox{0.9}{0.086}   &  \footnotesize{ F1IV-V                }   \\
\footnotesize{HD 173524}     &  \footnotesize{11323}  &    \scalebox{0.9}{3.93}  &    \scalebox{0.9}{0.10}    &  \scalebox{0.9}{0.015}   &  \footnotesize{ B9.5p...              }   \\
\footnotesize{HD 173667}     &  \footnotesize{6458}   &    \scalebox{0.9}{4.04}  &    \scalebox{0.9}{0.01}    &  \scalebox{0.9}{0.000}   &  \footnotesize{ F6V                   }   \\
\footnotesize{HD 194943}     &  \footnotesize{6971}   &    \scalebox{0.9}{4.04}  &    \scalebox{0.9}{-0.01}   &  \scalebox{0.9}{0.006}   &  \footnotesize{ F3V                   }   \\
\footnotesize{HD 205512$^\mathrm{c}$}     &  \footnotesize{4703}   &    \scalebox{0.9}{2.57}  &    \scalebox{0.9}{0.03}    &  \scalebox{0.9}{0.000}   &  \footnotesize{ K1III                 }   \\
\footnotesize{HD 206826}     &  \footnotesize{6490}   &    \scalebox{0.9}{4.09}  &    \scalebox{0.9}{-0.11}   &  \scalebox{0.9}{0.000}   &  \footnotesize{ F6V                   }   \\
\footnotesize{HD 207130}     &  \footnotesize{4741}   &    \scalebox{0.9}{2.65}  &    \scalebox{0.9}{0.08}    &  \scalebox{0.9}{0.010}   &  \footnotesize{ K0III                 }   \\
\footnotesize{HD 209459}     &  \footnotesize{11015}  &    \scalebox{0.9}{3.99}  &    \scalebox{0.9}{-0.07}   &  \scalebox{0.9}{0.052}   &  \footnotesize{ B9.5V                 }   \\
\footnotesize{NGC 7789 415}  &  \footnotesize{3815}   &    \scalebox{0.9}{1.16}  &    \scalebox{0.9}{0.01}    &  \scalebox{0.9}{0.269}   &  \footnotesize{ GB                    }   \\
\footnotesize{NGC 7789 501}  &  \footnotesize{4057}   &    \scalebox{0.9}{1.69}  &    \scalebox{0.9}{0.01}    &  \scalebox{0.9}{0.269}   &  \footnotesize{ GB                    }   \\
\footnotesize{NGC 7789 637}  &  \footnotesize{4857}   &    \scalebox{0.9}{2.54}  &    \scalebox{0.9}{0.01}    &  \scalebox{0.9}{0.269}   &  \footnotesize{ -                     }   \\
\footnotesize{NGC 7789 971}  &  \footnotesize{3746}   &    \scalebox{0.9}{1.22}  &    \scalebox{0.9}{0.01}    &  \scalebox{0.9}{0.269}   &  \footnotesize{ GB                    }   \\ 
    
    \hline
    \end{tabular}
    \begin{list}{Table Notes:}
    \item  \scriptsize{ \textbf{(1)} star identification; \textbf{(2)}  effective temperature; \textbf{(3)} superficial gravity; \textbf{(4)} metallicity; \textbf{(5)} extinction parameter and \textbf{(6)} spectral type. \textbf{(2)--(4)} were obtained from \citet{Prugniel+11,Sharma+16} and \textbf{(5)--(6)}, from \citet{Sanchez+06}. The superscript letter after the control star name identifies the library in which it is also present: $^\mathrm{a}$ in XSL; $^\mathrm{b}$ in IRTF; and $^\mathrm{c}$ in EIRTF.}
    \end{list}
    \label{tab:stars}
\end{table}

\subsection{Observations and data reduction} \label{sub:observation}

The NIR spectra were obtained using the cross-dispersed (XD) mode of GNIRS on the $8.1$~m Gemini North telescope in Mauna Kea, Hawaii. With the ``long blue'' camera with the LXD prism, $10$~l/mm grating and $0.10$\arcsec\ wide slit, this mode gives simultaneous spectral coverage from $\sim 0.83 - 2.5$~\micron\, at $R\sim1\,300$ with a pixel scale of $0.05$\arcsec/pix. To remove the sky emission, the targets have been observed in the ABBA-type pattern, with the source always on the slit. One telluric star per object was observed (either before or after the observations) to remove the telluric bands that plague the NIR spectral range. Individual and total exposure times varied depending on the object's brightness and the likely observing conditions (see below) and are given in Tab.~\ref{tab:obslog}.

\begin{table*}
\renewcommand{\tabcolsep}{1.5mm}
\renewcommand{\arraystretch}{0.9}
\centering
\caption{Observation log for the \textsc{smarty} stars.}
\begin{tabular}{cccccccccc}
\hline
\textbf{Star} & \textbf{RA}  & \textbf{DEC} & \textbf{Airmass} & \textbf{ExpTime} & \textbf{Observations} & \textbf{ImQuality} & \textbf{Cloud Cover} & \textbf{Background} & \textbf{Water Vapour}      \\
    \footnotesize{}& 
    \footnotesize{$^{\circ}$} & 
    \footnotesize{$^{\circ}$} & 
    \footnotesize{} &  
    \footnotesize{s} & 
    \footnotesize{\#} & 
    \footnotesize{\%-ile} & 
    \footnotesize{\%-ile} & 
    \footnotesize{\%-ile} & 
    \footnotesize{\%-ile}  \\    
    \textbf{\tiny{(1)}} & 
    \textbf{\tiny{(2)}} & 
    \textbf{\tiny{(3)}} & 
    \textbf{\tiny{(4)}} &  
    \textbf{\tiny{(5)}} & 
    \textbf{\tiny{(6)}} & 
    \textbf{\tiny{(7)}} & 
    \textbf{\tiny{(8)}} & 
    \textbf{\tiny{(9)}} & 
    \textbf{\tiny{(10)}}  \\    \hline
HD 015798       & 38.0218  & $-15.24467$ & 1.228            & 1.2              & 8                        & 70 & 50 & Any & 80  \\
HD 026322       & 62.7078  & 26.48095    & 1.073            & 2.7              & 12                       & 70 & 70 & 20  & 80  \\
HD 027295       & 64.8587  & 21.14231    & 1.293            & 5.5              & 12                       & 20 & 70 & Any & Any \\
HD 029375       & 69.5393  & 16.03329    & 1.019            & 4.0              & 12                       & 70 & 50 & 20  & 20  \\
HD 071030       & 126.4578 & 17.04627    & 1.143            & 4.3              & 12                       & 70 & 70 & Any & Any \\
HD 078234       & 137.0173 & 32.54040    & 1.125            & 5.7              & 12                       & 70 & 70 & Any & 20  \\
HD 087822       & 152.0662 & 31.60405    & 1.582            & 4.7              & 8                        & 70 & 50 & 50  & 50  \\
HD 113022       & 195.1613 & 18.37301    & 1.168            & 7.0              & 12                       & 70 & 50 & 50  & 80  \\
HD 114642       & 198.0148 & $-16.19860$ & 1.242            & 2.5              & 12                       & 70 & 70 & Any & 80  \\
HD 121299       & 208.6756 & $-1.50312$  & 1.321            & 1.1              & 12                       & 70 & 50 & Any & 50  \\
HD 137391       & 231.1226 & 37.37716    & 1.074            & 1.5              & 12                       & 70 & 50 & 50  & Any \\
HD 142908       & 238.9483 & 37.94696    & 1.159            & 3.5              & 12                       & 70 & 50 & 20  & Unknown\\
HD 143807       & 240.3607 & 29.85106    & 1.187            & 5.5              & 12                       & 70 & 50 & 80  & 80  \\
HD 145976       & 243.1895 & 26.67058    & 1.144            & 6.0              & 12                       & 70 & 50 & 80  & 80  \\
HD 149121       & 248.1487 & 5.52122     & 1.033            & 8.5              & 8                        & 70 & 70 & Any & 80  \\
HD 155078       & 257.4498 & $-10.52330$ & 1.321            & 3.7              & 12                       & 70 & 50 & Any & 80  \\
HD 157856       & 261.4911 & $-1.65178 $ & 1.108            & 8.5              & 12                       & 70 & 70 & 20  & Any \\
HD 166285       & 272.4751 & 3.11983     & 2.228            & 3.0              & 8                        & 20 & 70 & Any & Any \\
HD 169027       & 274.2447 & 68.74146    & 1.686            & 19.0             & 12                       & 20 & 70 & 50  & Any \\
HD 172103       & 279.5792 & $-1.11299$  & 1.944            & 7.0              & 12                       & 70 & 50 & Any & Any \\
HD 173524       & 280.6581 & 55.53946    & 1.732            & 3.7              & 12                       & 20 & 50 & Any & Any \\
HD 173667       & 281.4155 & 20.54631    & 1.842            & 1.0              & 8                        & 70 & 50 & Any & Any \\
HD 194943       & 307.2151 & $-17.81369$ & 1.570            & 1.2              & 12                       & 20 & 70 & Any & Any \\
HD 205512       & 323.6940 & 38.53406    & 1.177            & 0.8              & 8                        & 70 & 70 & 80  & Any \\
HD 206826       & 326.0358 & 28.74261    & 1.162            & 1.0              & 12                       & 20 & 70 & 80  & Any \\
HD 207130       & 325.7668 & 72.32009    & 1.685            & 1.0              & 4                        & 70 & 70 & 80  & 80  \\
HD 209459       & 330.8293 & 11.38655    & 1.133            & 7.5              & 12                       & 70 & 70 & 80  & Any \\
NGC 7789 415    & 359.2636 & 56.76609    & 1.530            & 13.5             & 12                       & 70 & 70 & Any & 50  \\
NGC 7789 501    & 359.2964 & 56.74142    & 1.404            & 45.0             & 12                       & 70 & 70 & Any & 20  \\
NGC 7789 637    & 359.3435 & 56.69611    & 1.253            & 120.0            & 4                        & 70 & 50 & Any & 50  \\
NGC 7789 971    & 359.4649 & 56.64907    & 1.402            & 15.0             & 12                       & 70 & 70 & Any & 50  \\ \hline
\end{tabular}
    \begin{list}{Table Notes:}
    \item  \footnotesize{ \textbf{(1)} star identification; \textbf{(2)} right ascension in degrees; \textbf{(3)} declination in degrees; \textbf{(4)} relative air mass; \textbf{(5)} exposure time \textbf{(6)} number of exposures \textbf{(7)} image quality in percentile; \textbf{(8)} cloud cover in percentile; \textbf{(9)} background in percentile when the information is provided; \textbf{(10)} water vapour in percentile when the information is provided. }
    \end{list}
    \label{tab:obslog}
\end{table*}

The slit was orientated close to the mean parallactic angle during the observations of both the science target and standard star. This procedure was adopted to minimise the effects of differential atmospheric refraction, which can be important over this wide wavelength range, especially at low elevations.

The data were acquired in queue mode between 2016 and 2017. Observations were taken from standard queue programs (GN-2016B-Q-76, GN-2017A-Q-66 -- PI: R. Riffel). 

Data reduction was carried out by a slightly modified version of XDGNIRS \citep{Mason+15} pipeline, V1.9, which is available at \url{ https://xdgnirs.readthedocs.io/en/latest/}.
Standard CCD procedures, such as bias subtraction, flat-fielding, and wavelength calibration, were followed and implemented via customary {\sc iraf} \citep{tody_iraf_1986} tasks.
Uncertainties were estimated from electron counts due to the science targets and atmospheric emission, as well as characteristic read noise and dark current values of the detector.

The two most critical aspects of the data reduction were removing telluric features and matching the sensitivity function between different orders.
Regarding the former, the reference spectrum of telluric standards was first treated with an algorithm to remove hydrogen lines from the star's atmosphere based on a direct comparison with a high-resolution and high signal-to-noise spectrum of Vega.
As for the latter, scale factors for small multiplicative corrections between the sensitivity functions of different diffraction orders were estimated by minimising the quadratic difference between overlapping regions of the spectrum. The same telluric standard star was used for flux calibration.

\section{Data quality} \label{sec:results}

Many of the {\sc smarty} observations at Gemini were carried out under poor weather conditions (see Tab.~\ref{tab:obslog}). Despite that, we achieved a good telluric correction, but we needed to apply an independent flux calibration to our data in order to correct the 'steps' in the continuum. These irregularities likely stem from challenges to match the sensitivity function of different orders. The independent flux calibration was performed according to the following procedures. First, we normalised the {\sc smarty} spectra, leading to a spectrum of pure absorption features, $F_{\rm norm}$. We then multiplied the normalised spectra by the continuum flux, $F_{\rm C}$, from a reference spectra, using:

\begin{enumerate}
    \item {\bf common stars}: Five {\sc smarty} stars have NIR spectra available from other empirical libraries, namely:  HD\,027295 ($T_\mathrm{eff}=11\,034$\,K; in XSL), HD\,087822 ($T_\mathrm{eff}=6\,573$\,K; in IRTF), HD\,137391, HD\,145976 and HD\,205512 ($T_\mathrm{eff}=7\,186$, $6\,927$ and $4\,703$\,K, respectively; in EIRTF). The shape of the continuum of these stars has been used as a reference to fine-tune the flux calibration of the {\sc smarty} counterpart. Note that these stars have a good $T_\mathrm{eff}$ coverage.
   
    \item {\bf interpolated stars}: For all {\sc smarty} stars, we have done the independent flux calibration using the {\sc e-miles} interpolator \citep[see][for more details]{Vazdekis+03,Vazdekis+16} to interpolate among 180 IRTF plus 200 EIRTF stellar spectra to compute a spectrum which best matches the {\sc smarty} stars stellar parameters adopting a local interpolation scheme\footnote{The interpolator selects stars whose parameters are within a box around the requested parametric point ($T_\mathrm{eff}$, $\log{g}$, [Fe/H]), which is divided into eight boxes. If no stars are found in any of these boxes, it can be expanded up to a limit. Thus, the larger the density of stars around the point, the smaller the box is.}. Therefore, as IRTF+EIRTF do not have a significant number of stars hotter than $\sim7\,000$\,K, the interpolated corrected flux is not recommended for stars above this temperature since the box can be bigger than typical uncertainties in the determination of the parameters.

    \item {\bf synthetic stars}: Following \citet{Coelho+20}, we computed synthetic spectra for the 31 {\sc smarty} stars using as input the values computed by \citet{Prugniel+11} and \citet{Sharma+16} for effective temperature and surface gravity, and approximated values of metallicity ([Fe/H]\,$=-0.1$, $0.0$ or $0.2$). The continuum of these stars has been used as a reference for the independent flux calibration. It is worth noting that the spectra of colder ($T_\mathrm{eff}<6\,000$\,K) are not well predicted by the models\footnote{For the optical wavelengths, \citet{Coelho+14} has shown that the flux differences between the synthetic and the \textsc{miles} similar parameters stars are within 2\% for stars with $T_\mathrm{eff}\geq6\,250$\,K, and 5\% for stars with $T_\mathrm{eff}\geq4\,750$\,K, but can reach 50\% for colder stars.}.
\end{enumerate}

\begin{figure*}
    \centering
    \begin{tabular}{ccc}
    \includegraphics[width=0.99\hsize,page=1]{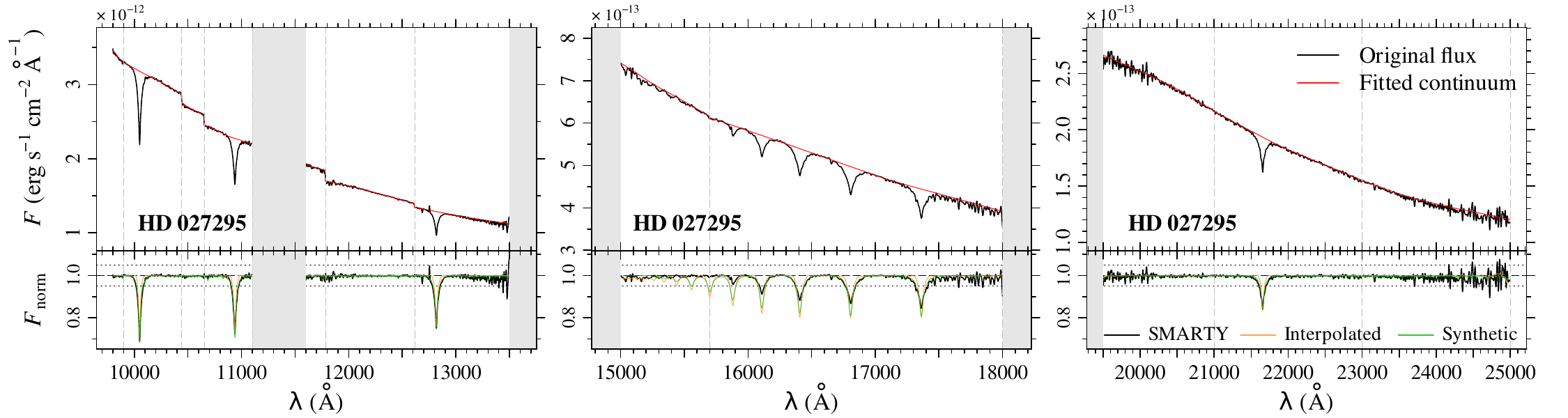}  \\

    \includegraphics[width=0.99\hsize,page=1]{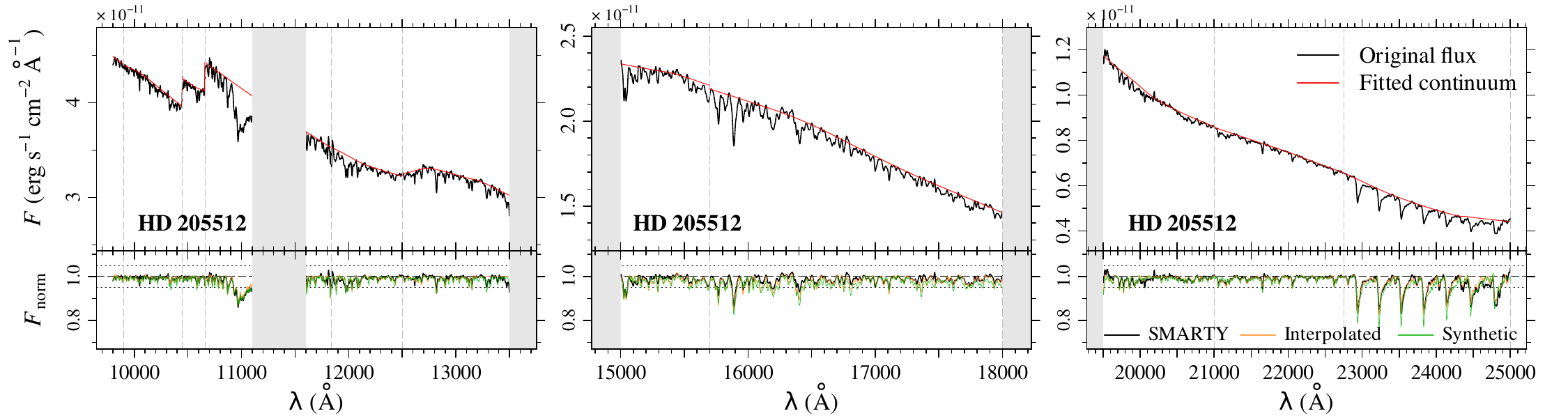}  \\
    \end{tabular}
    
    \caption{Illustration of the continuum-fit process for a hot (HD\,27295, $T_{\rm eff} \sim 11\,000$\,K) and a cold star (HD\,205512, $T_{\rm eff} \sim 4\,700$\,K). The upper plot in each panel shows the original flux (\emph{black line}) and the fitted continuum (\emph{red line}), and the lower plot shows the resulting normalised flux. The normalised spectra of the synthetic and interpolated stars in the lower plots are also shown as the \emph{orange} and \emph{green lines}, respectively. To properly fit the continuum of different spectral regions and to correct the `steps' in the flux that could not be removed during the data reduction, the fit was done independently within wavelength ranges indicated by the vertical \emph{grey dashed lines} (see text for details). The figures showing the fitted continuum for all the stars are in the Appendix~\ref{Ap:cont_fit} (Fig.\ref{fig:cont_fit1}).}
    \label{fig:cont_fit}
\end{figure*}

To fit the continuum, we smoothed the spectra using LOESS\footnote{Locally Estimated Scatterplot Smoothing: a non-parametric method for local polynomial regression. We used the task {\tt loess} from the {\tt stats} R package \citep{R:2021}.}, discarding the low flux values (i.e., the absorption features) from the spectra in each iteration by adopting different values for the lower and upper $\sigma$-clipping factors. This process was repeated until only the data points of the continuum were left to be fitted. To properly fit the continuum of different spectral regions, the fit was done independently within wavelength ranges, as illustrated in Fig.~\ref{fig:cont_fit}. The whole procedure was performed interactively by visually inspecting the fits and changing the fitting parameters for each spectrum to achieve a good continuum fit. The parameters that can be adjusted in our approach are the lower and upper $\sigma$-clipping factors, the number of $\sigma$-clipping iterations, and the LOESS smoothing parameter, $\alpha$, which corresponds to the fraction of total number of data points that are used in each local fit. 

\begin{figure}
    \centering
    \includegraphics[width=0.95\hsize]{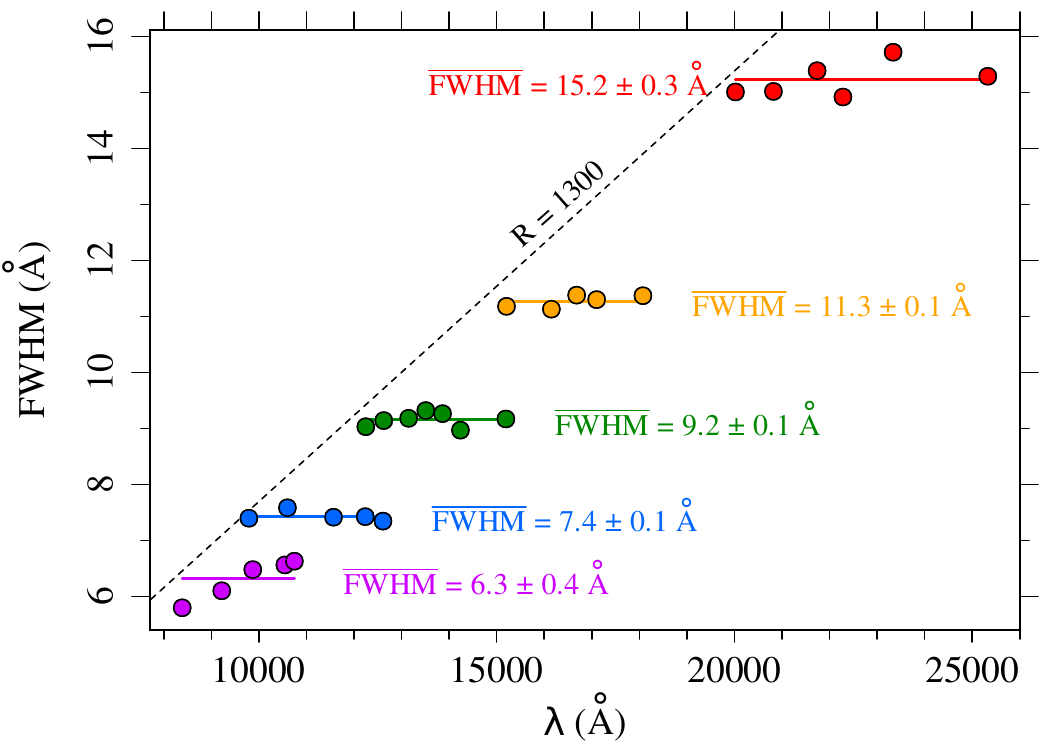}
    \caption{GNIRS spectral resolution as a function of wavelength, which is assumed to be the initial resolution of {\sc smarty} spectra. The {\sc smarty} spectra are convolved to constant $R = 1\,300$, indicated by the dashed line.}
    \label{fig:gnirs_res}
\end{figure}

In Fig.~\ref{fig:cont_fit}, we illustrate the continuum fitting process for a hot (HD\,27295, $T_{\rm eff} \sim 11\,000$\,K) and a cold star (HD\,205512, $T_{\rm eff} \sim 4\,700$\,K) as examples; similar figures for the other {\sc smarty} stars are shown in Appendix~\ref{Ap:cont_fit}. The same approach was adopted to fit the continuum of the reference stars to obtain $F_{\rm C}$. Before fitting the continuum, all spectra were degraded to the same resolution of $R = 1\,300$. The initial resolution of {\sc smarty} spectra, shown in Fig.~\ref{fig:gnirs_res}, was obtained from the arc lamp spectra.

The continuum-corrected {\sc smarty} spectra were multiplied by a factor so that the total flux within the $J$, $H$, and $K$ bands ($F_{\rm J}$, $F_{\rm H}$, and $F_{\rm H}$) of our final spectra is consistent with that of the 2MASS photometry; i.e., the factor was chosen so that it leads to $(F_{\rm J} + F_{\rm H} + F_{\rm K})_{\rm SMARTY} = (F_{\rm J} + F_{\rm H} + F_{\rm K})_{\rm 2MASS}$.

The line-of-sight velocities used to correct the {\sc smarty} spectra were determined through cross-correlation with the theoretical spectra using the task {\tt xcsao} from the {\tt iraf rvsao} package \citep[for details, see][]{Tonry+79,Kurtz+92,Mink+98}. To avoid the regions with telluric contamination, the cross-correlation was performed separately within the \textit{J}, \textit{H}, and \textit{K} wavelength ranges, and we adopted the result with the lowest uncertainty.

The flux-corrected \textsc{smarty} spectra are available at \url{www.if.ufrgs.br/~riffel/smarty/}. 
We recommend using {\sc smarty} with the continuum from the common stars when available, from interpolated stars when $T_{\rm eff} \lesssim 7500\,$K, and from synthetic stars when $T_{\rm eff} \gtrsim 7500\,$K.

\subsection{Comparison with literature data}
\begin{figure*}
    \centering
    \begin{tabular}{ccc}
    \includegraphics[width=0.99\hsize,page=1]{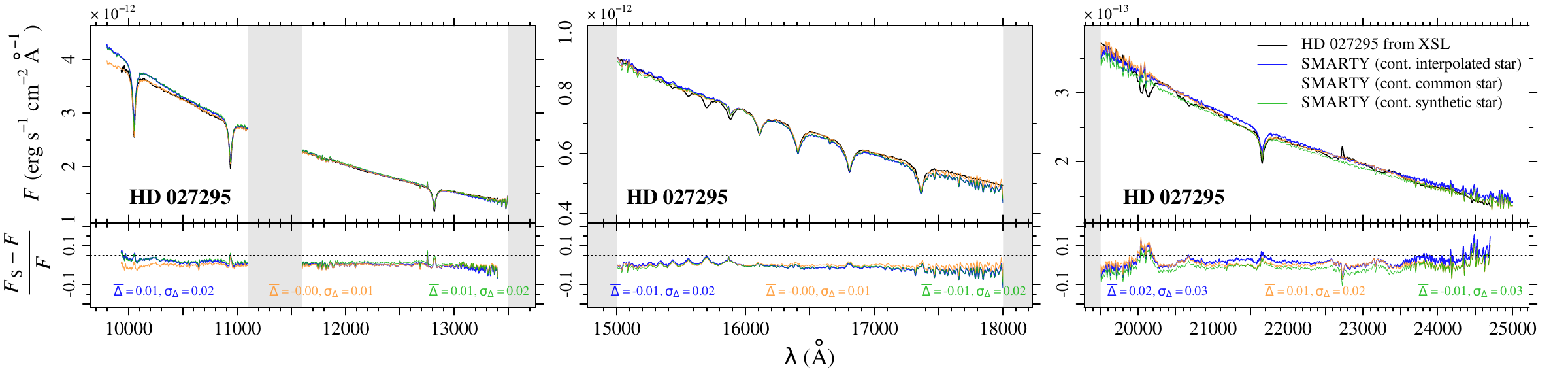}  \\

    \includegraphics[width=0.99\hsize,page=1]{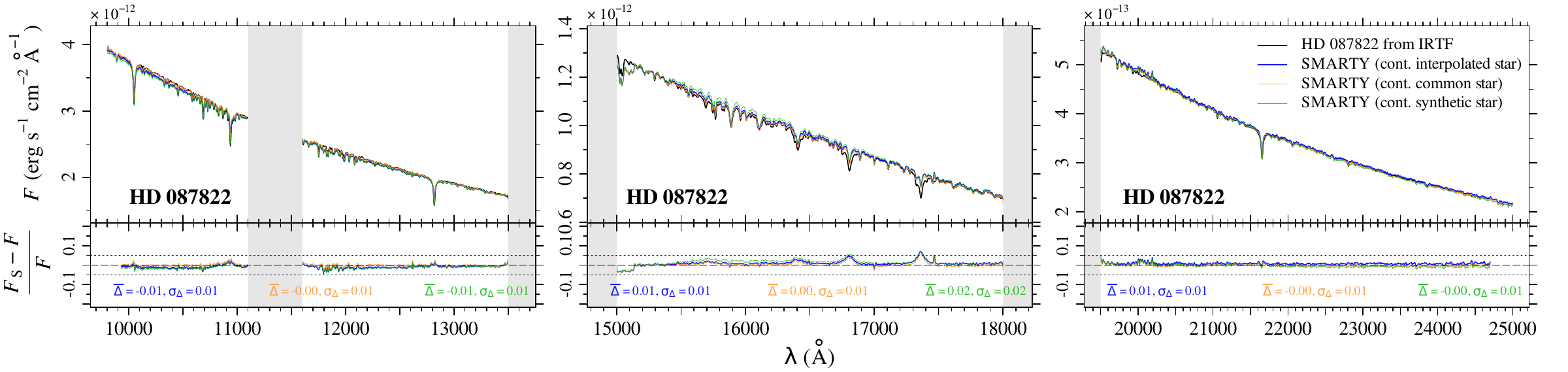}  \\

    \includegraphics[width=0.99\hsize,page=1]{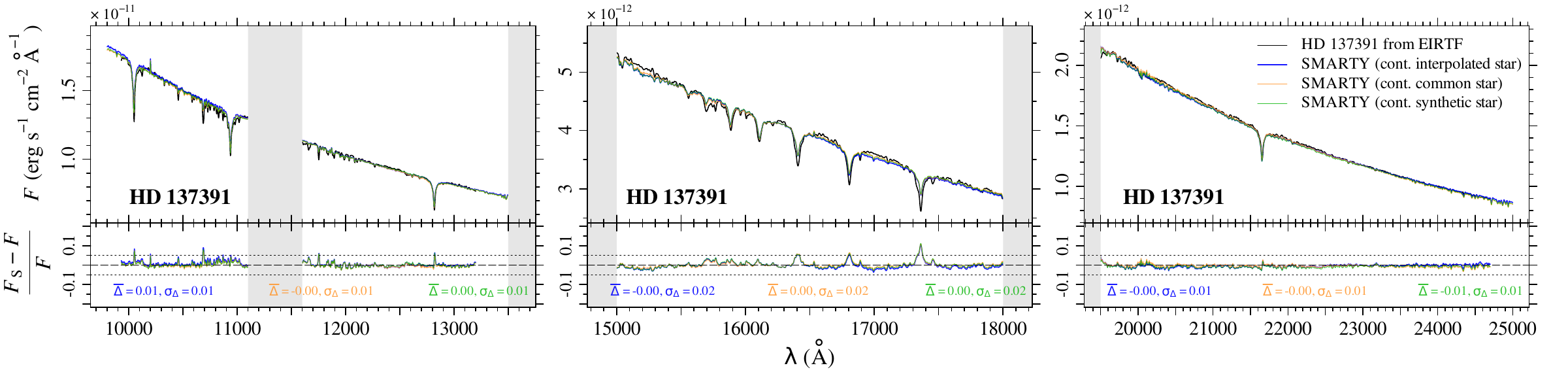}  \\

    \includegraphics[width=0.99\hsize,page=1]{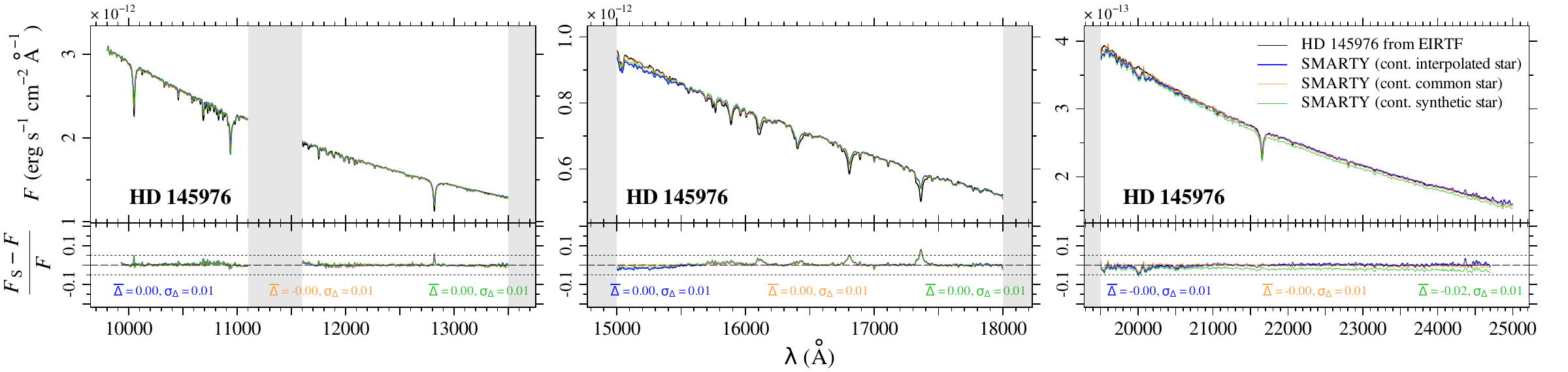}  \\

    \includegraphics[width=0.99\hsize,page=1]{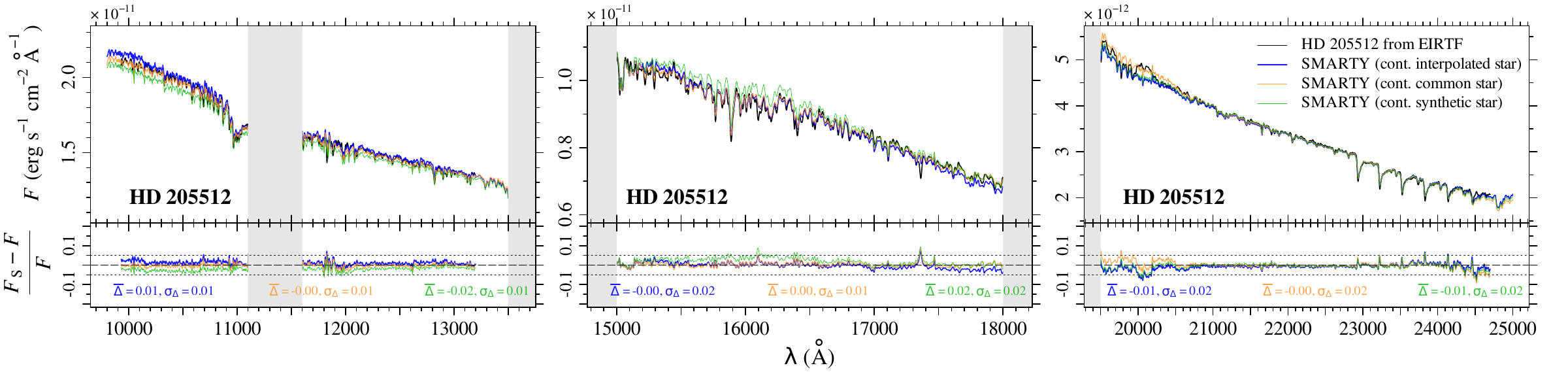}  \\

    \end{tabular}
    
    \caption{Comparison between \textsc{smarty} and the reference spectra from XSL, IRTF or EIRTF for the five stars in common with these libraries. The upper plot in each panel is the flux, and the lower plot shows the relative difference between the \textsc{smarty} and reference spectra. The dashed and dotted lines indicate $0 \pm 0.05$, respectively. The shaded areas are omitted due to telluric contamination.}
    \label{fig:common_zoom}
\end{figure*}

To assess the uncertainties of our approach, in Fig.~\ref{fig:common_zoom}, we show the {\sc smarty} spectra calibrated using $F_{\rm C}$ from the common, interpolated and synthetic stars (as listed above). In this figure, we show the five stars that also have spectra available in other NIR libraries, and to evaluate possible differences in a more quantitative way, we computed the pixel deviation following the equation:
\begin{equation}
   \Delta = \frac{(F_{S} - F) }{F},
\end{equation}
where $F_S$ is the flux of the {\sc smarty} star (for each of the independent calibration procedures), and $F$ is the flux of the common star (taken from the different libraries). We show the mean value of the pixel deviation ($\bar{\Delta}$) and its standard deviation ($\sigma_{\Delta}$) for the three independent flux calibration procedures in the different windows. 
As can be seen, there is a very good agreement for all 5 stars, with differences up to $\bar{\Delta}$ = 0.03 (for $H$-band of HD\,087822), but for almost all the cases, $\bar{\Delta} \sim$  0.01.

\subsection{Index comparisons}

The equivalent widths (EW) of absorption lines are one of the most adequate ways to compare the underlying spectrum of stars. To better compare the {\sc smarty} spectra with the predictions of their synthetic and interpolated versions, we have measured the EW of the ions of \ion{H}{i}, \ion{Mg}{i}, \ion{Fe}{i}, \ion{Mn}{i}, \ion{Al}{i}, \ion{Si}{i}, \ion{Na}{i}, \ion{Ca}{i} and molecular absorptions of CN and CO. For this, we have used a Python modified version of the code {\sc pacce} \citep{Riffel+11c} with the definitions for the indices presented in \citep{Riffel+19}, except for the \ion{Pa}{$\beta$} and \ion{Br}{$\gamma$} indices, where the definitions of \citet{Eftekhari+21} and \citet{Kleinmann+86} have respectively been used. Before measuring the EW, all the spectra have been homogenised to a uniform spectral resolution of $R=1\,300$ (see Fig.~\ref{fig:gnirs_res}).

\begin{figure*}
    \centering
    \includegraphics[width=0.94\hsize]{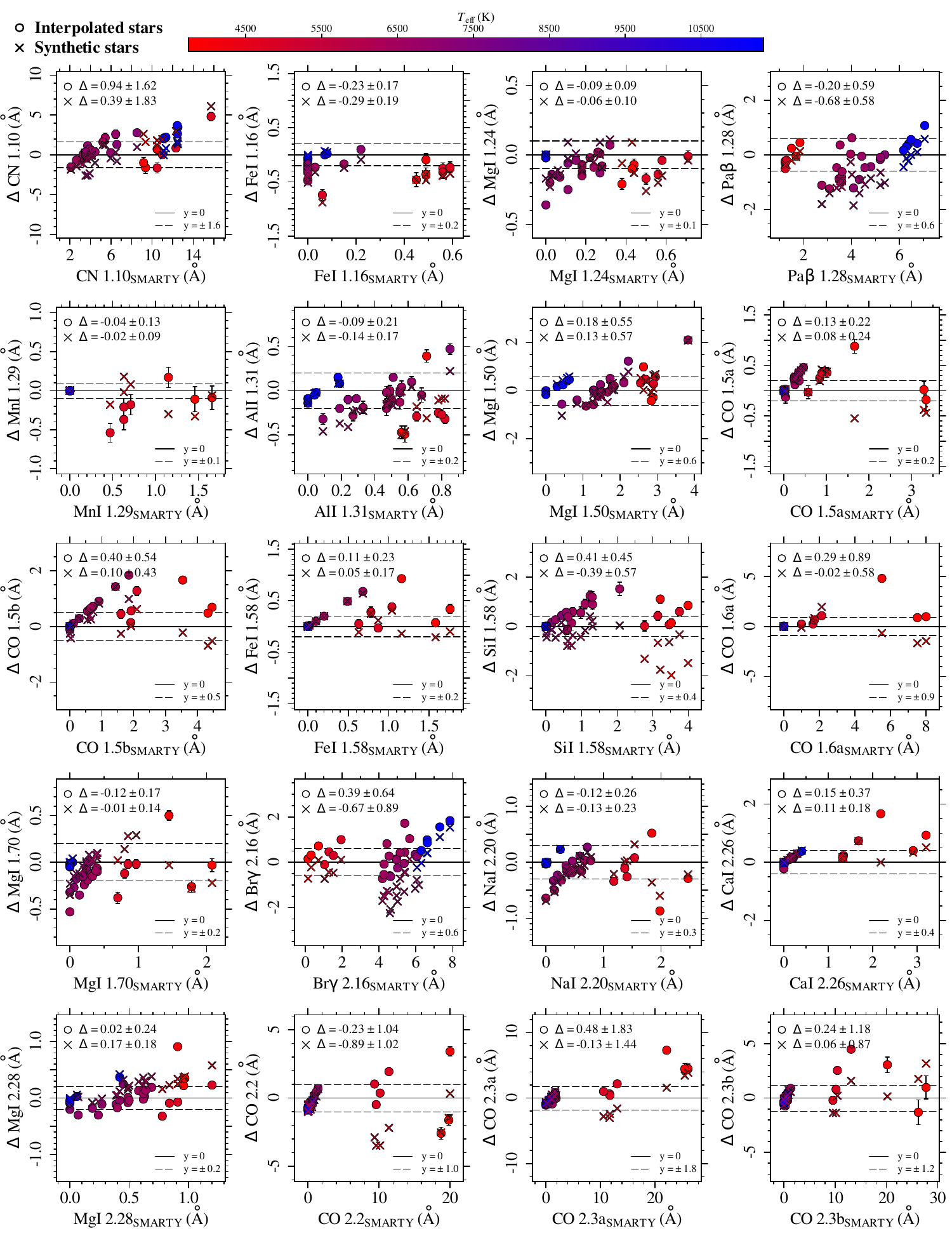}
    \caption{Differences between spectral indices measured in {\sc smarty} and the interpolated (\emph{circles}) and the synthetic (\emph{crosses}) stellar spectra vs. values obtained for {\sc smarty}. The symbols are colour-coded by effective temperature, and the error bars are shown for the measurements made in {\sc smarty} spectra with the continuum from interpolated stars. In each panel, for each set of measurements, we show the mean and standard deviation of the differences between {\sc smarty} and interpolated and synthetic measurements ($\Delta\pm\sigma_\Delta$).   The solid and the dashed lines represent $y = 0$ and $y = \pm \sigma_\Delta$, where $\sigma_\Delta$ is the standard deviation of the differences between the {\sc smarty} spectral indices and those from interpolated stellar spectra.}
    \label{fig:indices}
\end{figure*}

In Fig.~\ref{fig:indices}, we compare the indices measured in the {\sc smarty} with the predicted values of the interpolated and synthetic spectra. We show the measured value, for each index, in the {\sc smarty} stars in the x-axis and the difference between the {\sc smarty} stars with those measured ($\Delta$) in the synthetic spectra (crosses) and the interpolated ones (circles) in the y-axis. The {\sc smarty} measurements are colour-coded according to the star's temperature. The one-to-one relation is represented by the full line, while the standard deviation of the difference between the measurements of the {\sc smarty} and synthetic stars is represented as a dashed line.  

\begin{figure*}
    \centering
    \includegraphics[width=0.98\hsize]{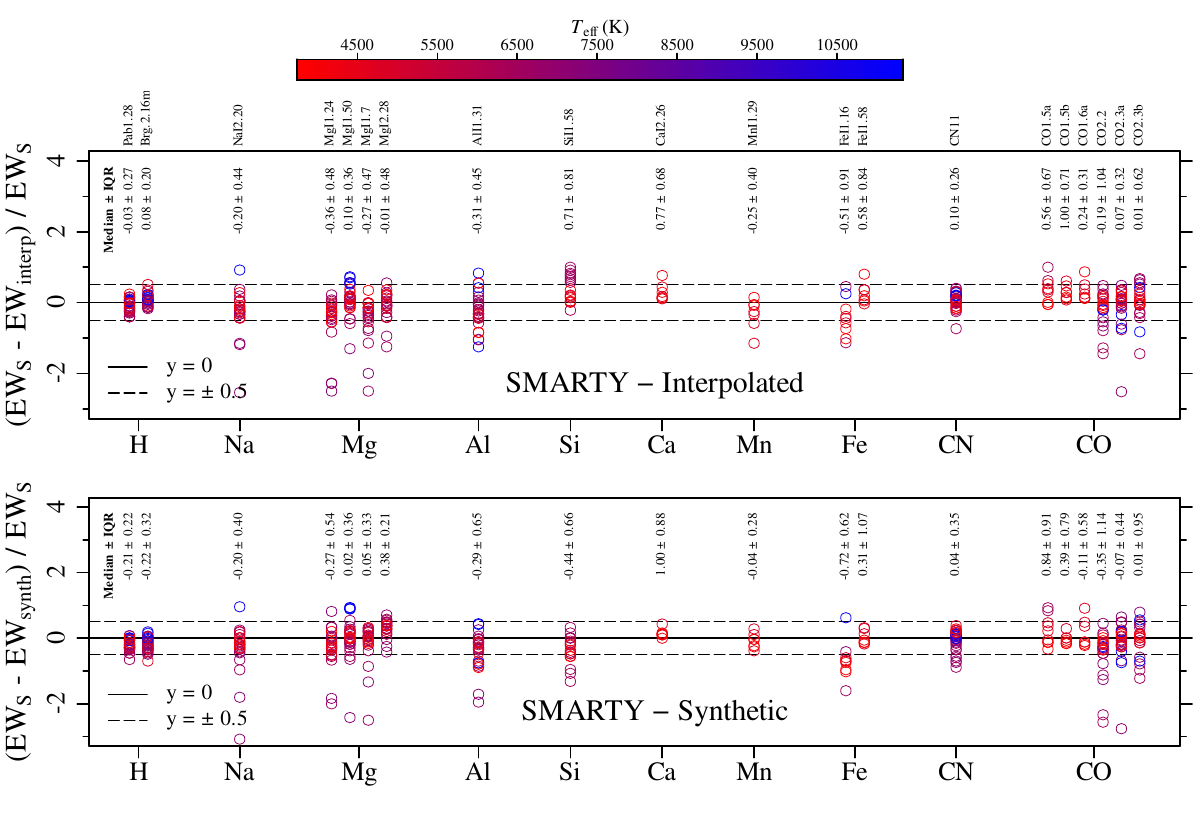}
    \caption{Relative differences between {\sc smarty} spectral indices and those measured in the interpolated (\emph{upper panel}) and the synthetic (\emph{lower panel}) stellar spectra, grouped by chemical element or molecule. In both panels, the symbols are colour-coded by effective temperature. The figure shows only stars for which ${\rm EW}/\sigma_{\rm EW} \geq 2$, where ${\rm EW}$ and $\sigma_{\rm EW}$ are the {\sc smarty} index value and its uncertainty, respectively. The median values and interquartile ranges (IQR) of the relative differences are shown for each index.}
    \label{fig:indices_type}
\end{figure*}

In general, there is a good agreement between the \textsc{smarty} and other (synthetic and interpolated) values. However, a large spread is observed among different indices. 
Although for some indices, the standard deviation of the differences (indicated by the dashed lines in Fig.~\ref{fig:indices}) is small, measurements for individual elements in individual stars may show very large differences (see Fig.~\ref{fig:indices_type}).
For instance, the indices \ion{Mn}{i}, \ion{H}{i} (especially \ion{Pa}{$\beta$}), and CN are similar for both the \textsc{smarty} spectra corrected by interpolated and by synthetic stars, with relative differences within 5\%. The worst indices were the CO bands, \ion{Si}{i} and \ion{Ca}{i}, reaching up to a relative difference of up to 2 times. Also, in some cases, the difference between the synthetic and interpolated measurements is too large (e.g. \ion{Si}{i}\,$\lambda\,15\,800$\,\AA), where the interpolated values are smaller than the synthetic ones, which can reach up to 2\,\AA\ larger than the interpolated ones for the cooler stars.

Taking into account all the indices at the same time, we found that $({\rm EW}_\mathrm{S} - {\rm EW}_\mathrm{synth})/{\rm EW}_\mathrm{S} = -0.2\pm0.84$ and $({\rm EW}_\mathrm{S} - {\rm EW}_\mathrm{interp})/{\rm EW}_\mathrm{S} = -0.13\pm0.81$, indicating that, within the errors, both interpolated and synthetic spectra are in good agreement, with the interpolated EW predictions being somehow in better agreement than those obtained from the synthetic ones.

Finally, the source of these discrepancies is unclear, and addressing them is out of the scope of this study. However, we speculate that regarding the synthetic spectra, the spread may come from differences in the chemical abundance pattern (the synthetic spectra adopt solar-scaled abundances) or inaccuracies in the model opacities \citep[see][]{Coelho+14,Coelho+20}. In the case of the interpolated stars, the spectrum is computed by mixing observed spectra of different stars \citep[see][]{Vazdekis+03}, thus in terms of individual elements, it may produce some deviations (e.g. differences in elemental abundances may produce different indices).

\subsection{Comparison with 2MASS photometry} \label{sec:photometry}

To check the overall accuracy of the final corrected \textsc{smarty} spectra\footnote{We compared the magnitudes from the flux second order calibration of the \textsc{smarty} stars, for both corrections, e.g. using the continuum from the interpolated and the theoretical spectra.}, we compared the spectrophotometric magnitudes derived from the \textsc{smarty} spectra with the values available at the 2MASS point source catalogue \citep[PSC,][]{Skrutskie+06}. 

\begin{figure*}
    \centering
    \includegraphics[width=0.9\textwidth]{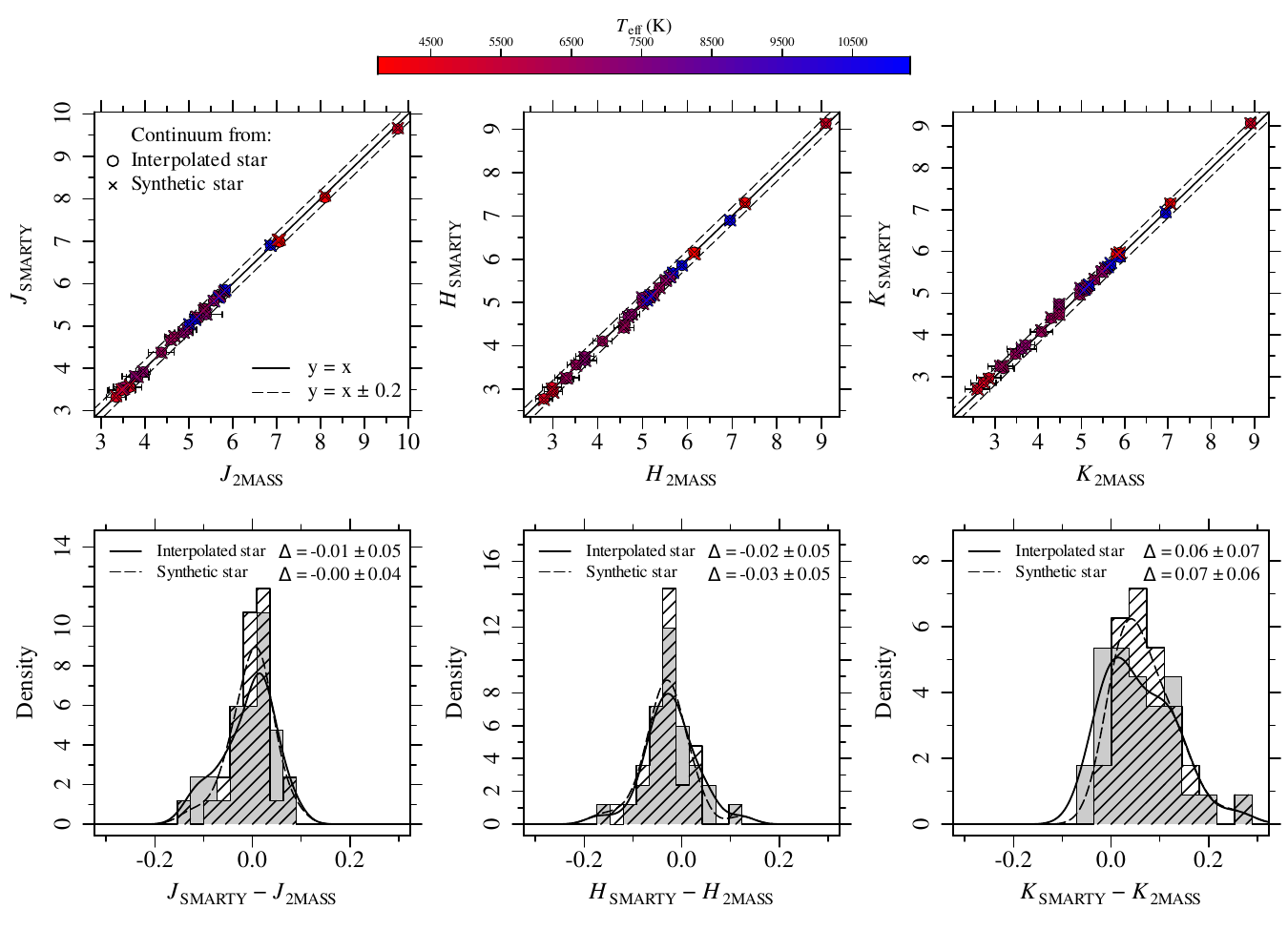}
    \caption{Comparison between the 2MASS magnitudes with those from the \textsc{smarty} spectra corrected using as the continuum from the interpolated (\emph{circles}) and theoretical stars (\emph{crosses}), colour-coded by effective temperature. 
    \textbf{Upper panels:} \textit{J}, \textit{H}, and \textit{K} magnitudes from the \textsc{smarty} spectra vs. 2MASS magnitudes. The \emph{black solid} and \emph{dashed lines} indicate $y = x$ and $y = x \pm 0.2$, respectively. \textbf{Bottom panels:} distributions of the differences between the 2MASS and \textsc{smarty} magnitudes obtained with \textsc{smarty} spectra corrected using the interpolated (\emph{filled grey histograms}) and theoretical stars (\emph{black hashed histograms}). The distribution mean values and standard deviations are indicated in each panel. The \emph{solid} and the \emph{dashed lines} are obtained by smoothing the positions of the data points using a Gaussian kernel with a standard deviation equal to half of the standard deviation of the data points.}
    \label{fig:photometry}
\end{figure*}

In Fig. \ref{fig:photometry}, we compare \textsc{smarty} and 2MASS magnitudes in $J$, $H$ and $K$ bands. We can see that there is a very good agreement between \textsc{smarty} and 2MASS magnitudes, with maximum differences that do not exceed $\sim0.2$\,mag. The mean differences in magnitude for the stars corrected by interpolated and synthetic spectra are, respectively, $-0.01$ and $0.0$\,mag for the $J$ band; $-0.02$ and $-0.03$\,mag for the $H$ band; and $0.06$ and $0.07$\,mag for the $K$ band. 

\begin{figure*}
    \centering
    \includegraphics[width=0.9\textwidth]{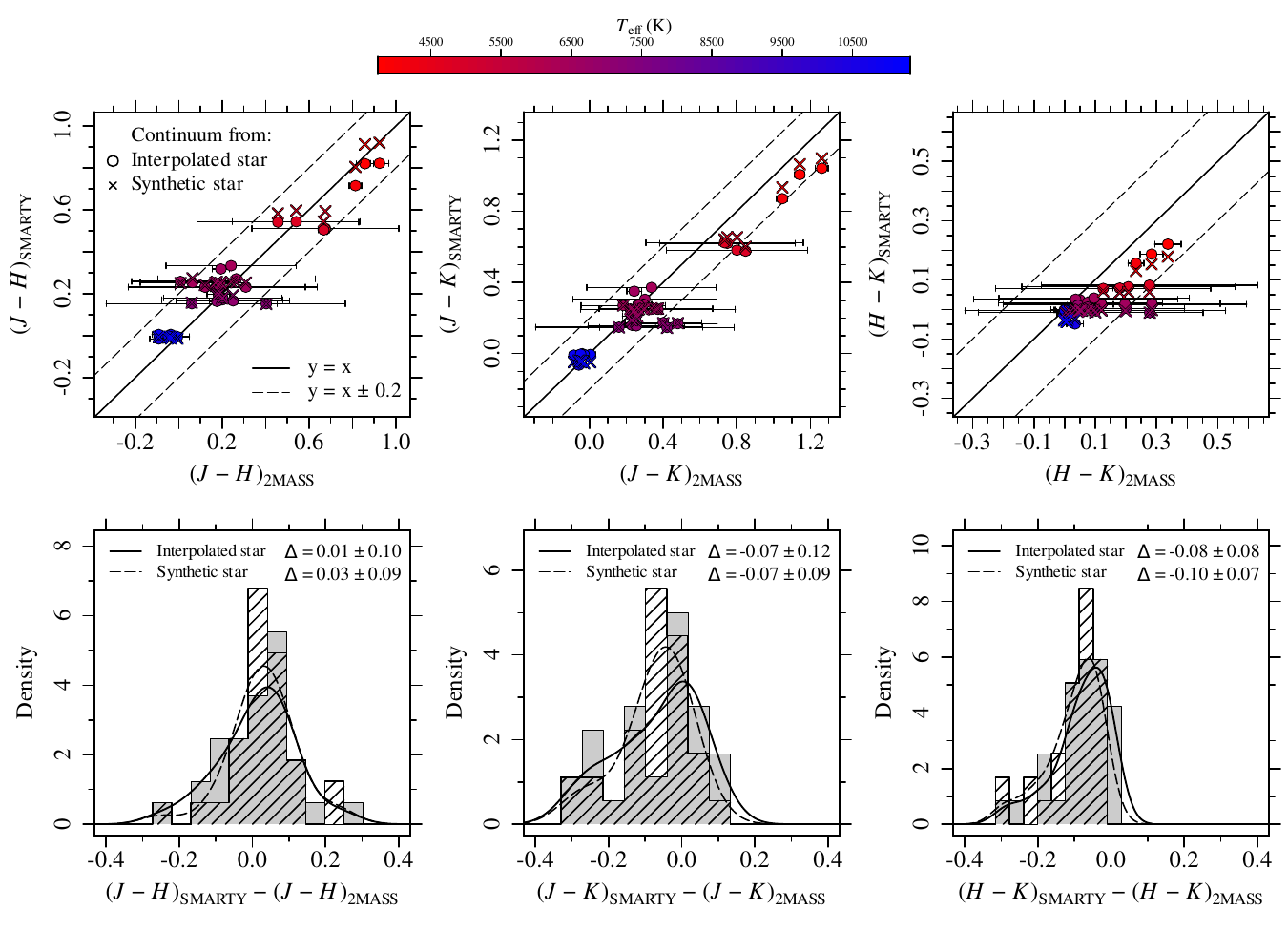}
    \caption{Comparison between the 2MASS colour indices with those from the \textsc{smarty} spectra corrected using as the continuum from the interpolated (\emph{circles}) and theoretical stars (\emph{crosses}), colour-coded by effective temperature.
    \textbf{Upper panels:} $(J-H)$, $(J-K)$, and $(H-K)$ from the \textsc{smarty} spectra vs. 2MASS colour indices. The \emph{black solid} and \emph{dashed lines} indicate $y = x$ and $y = x \pm 0.2$, respectively. \textbf{Bottom panels:} distributions of the differences between the 2MASS and \textsc{smarty} magnitudes obtained with \textsc{smarty} spectra corrected using the interpolated (\emph{filled grey histograms}) and theoretical stars (\emph{black hashed histograms}). The distribution mean values and standard deviations are indicated in each panel. The \emph{solid} and the \emph{dashed lines} are obtained by smoothing the positions of the data points using a Gaussian kernel with a standard deviation equal to half of the standard deviation of the data points.}
       \label{fig:photometry_color}
\end{figure*}

The colour indices obtained directly from the {\sc smarty} spectra are compared with the 2MASS ones in Fig.~\ref{fig:photometry_color}, where a good agreement can be observed with the maximum differences being smaller than $\sim 0.3$\,mag for all colour indices. However, the \textsc{smarty} spectra lead to slightly lower (bluer) colour indices compared to the values from the 2MASS photometry (with mean differences in the range of $-0.10$ to $-0.07$\,mag in both the $J-K$ and $H-K$ indices). Part of this small systematic difference might be due to Galactic extinction corrections, which are not applied to the 2MASS magnitudes, but, since we use the continuum from the reference stars, the \textsc{smarty} spectra are "implicitly" extinction corrected. 

To estimate the impact of the extinction corrections on the colour comparisons, we computed the colour differences for stars with $E(B-V) \simeq 0$ only; we find that the differences between {\sc smarty} and 2MASS colours are smaller for these stars, with $\Delta(J-K) = 0.00 \pm 0.09$ ($-0.03 \pm 0.07$) and $\Delta(H-K) = -0.06 \pm 0.06$ ($-0.09 \pm 0.05$) for {\sc smarty} stars with continuum corrected using the interpolated (synthetic) stars. On the other hand, $\Delta(J-H)$ increases to $0.06 \pm 0.07$ ($0.05 \pm 0.07$).  We also obtained extinction estimates in the $J$, $H$, and $K$ bands by \citet{Schlegel+98}\footnote{Extinction estimated were obtained through the NASA/IPAC Infrared Science Archive (IRSA) web tool available at \url{https://irsa.ipac.caltech.edu/applications/DUST/}. These estimates correspond to the total reddening along the line of sight, and the distances to the stars are not considered.} and corrected the 2MASS magnitudes and colours. After applying the corrections, the differences between {\sc smarty} and 2MASS colours decrease to $\Delta(J-K) = 0.04 \pm 0.14$ ($0.04 \pm 0.13$) and $\Delta(H-K) = -0.04 \pm 0.08$ ($-0.06 \pm 0.07$). However, $\Delta(J-H)$ increases to $0.09 \pm 0.12$ ($0.10 \pm 0.12$).

\section{Final Remarks} \label{sec:conclusions}

We presented a stellar spectral library with 31 stars covering the wavelength range from $0.9$ to $2.4\,$\micron\, observed with GNIRS at Gemini North Telescope. The \textsc{smarty} is publicly available at \url{www.if.ufrgs.br/~riffel/smarty/}. To ensure the spectra quality, we corrected the flux using the continuum of reference stellar spectra from three different sources: $i$) stars in common with other NIR empirical libraries; $ii$) spectra obtained through interpolation of the empirical IRTF+EIRTF library; and $iii$) theoretical spectra based on \citet{Coelho+20}, extended to cover our wavelength range.

The average flux difference between \textsc{smarty} and the reference spectra is $\lesssim 2\%$, as can be seen in Fig.~\ref{fig:common_zoom}, where we compare the \textsc{smarty} spectra corrected with the continuum from the three sources mentioned above with the spectra of stars in common with other libraries. 

We also investigated our data reliability by comparing the equivalent widths measured in the {\sc smarty} spectra with those obtained from synthetic spectra computed with the atmospheric parameters of the {\sc smarty} stars and interpolated from the IRTF+EIRTF stars. We find good agreement between the EW values; however, large differences can be observed for a few individual stars.

We have also compared the magnitudes and colour indices from \textsc{smarty} spectra with those from 2MASS photometry. The comparison reveals a very good agreement between \textsc{smarty} and 2MASS magnitudes, with mean differences from $-0.01$ to $0.07$\,mag and standard deviation from $0.04$ to $0.07$\,mag in the $J$, $H$, and $K$ bands. A good agreement is also observed for the colour indices, with mean differences from $-0.10$ to $0.03$\,mag for the $(J-H)$, $(J-K)$, and $(H-K)$ indices.  A small difference was noted between the \textsc{smarty} spectra corrected using the continuum from the interpolated and the theoretical stars.




\section*{Acknowledgements}

We thank the anonymous referee for the valuable feedback, which greatly improved this manuscript.
We thank Alan Alves Brito and Alejandra Romero for their helpful comments and discussions. 
MBC acknowledges support from Funda\c{c}\~ao de Amparo \`{a} Pesquisa do Rio Grande do Sul (FAPERGS) and Coordena\c{c}\~ao de Aperfei\c{c}oamento de Pessoal de N\'{i}vel Superior (CAPES). RR acknowledges support from the Fundaci\'on Jes\'us Serra and the Instituto de Astrof{\'{i}}sica de Canarias under the Visiting Researcher Programme 2023-2025 agreed between both institutions. RR also acknowledges support from the ACIISI, Consejer{\'{i}}a de Econom{\'{i}}a, Conocimiento y Empleo del Gobierno de Canarias and the European Regional Development Fund (ERDF) under grant with reference ProID2021010079, and the support through the RAVET project by the grant PID2019-107427GB-C32 from the Spanish Ministry of Science, Innovation and Universities MCIU. This work has also been supported through the IAC project TRACES, which is partially supported through the state budget and the regional budget of the Consejer{\'{i}}a de Econom{\'{i}}a, Industria, Comercio y Conocimiento of the Canary Islands Autonomous Community. RR also thanks to Conselho Nacional de Desenvolvimento Cient\'{i}fico e Tecnol\'ogico (CNPq, Proj. 311223/2020-6, 304927/2017-1 and 400352/2016-8), FAPERGS (Proj. 16/2551-0000251-7 and 19/1750-2), and CAPES (Proj. 0001). MT thanks the support from CNPq (process 312541/2021-0). LGDH acknowledges support by National Key R\&D Program of China No.2022YFF0503402. PC acknowledges support from CNPq under grant 310555/2021-3 and from Funda\c{c}\~{a}o de Amparo \`{a} Pesquisa do Estado de S\~{a}o Paulo (FAPESP) process number 2021/08813-7. DRD acknowledges support from CNPq under grant 313040/2022-2. AV acknowledges support from grant PID2021-123313NA-I00 and PID2022-140869NB-I00 from the Spanish Ministry of Science and Innovation.

This publication makes use of data products from the Two Micron All Sky Survey, which is a joint project of the University of Massachusetts and the Infrared Processing and Analysis Center/California Institute of Technology, funded by the National Aeronautics and Space Administration and the National Science Foundation.

\section*{Data Availability} \label{sec:data_availability}

The \textsc{smarty} presented in this article is available at \url{https://www.if.ufrgs.br/~riffel/smarty/}. We provide the following information in a single \texttt{csv} file for each star:

\begin{enumerate}
    \item \texttt{lambda}: wavelength (\AA);
    \item \texttt{flux}: rest-frame flux of the \textsc{smarty} spectrum corrected using the continuum of the interpolated star (erg\,cm$^{-2}$\,s$^{-1}$\,\AA$^{-1}$); 
    \item \texttt{err}: error on the flux corrected using the interpolated star (erg\,cm$^{-2}$\,s$^{-1}$\,\AA$^{-1}$);
    \item \texttt{flux\_norm}: normalized \textsc{smarty} spectrum;
    \item \texttt{err\_norm}: error on the \textsc{smarty} normalized spectrum;
    \item \texttt{flux\_cont\_interp\_star}: the continuum of the interpolated star (erg\,cm$^{-2}$\,s$^{-1}$\,\AA$^{-1}$);
    \item \texttt{flux\_corr\_synt}: rest-frame flux of the \textsc{smarty} spectrum corrected using the continuum of the synthetic star (erg\,cm$^{-2}$\,s$^{-1}$\,\AA$^{-1}$);
    \item \texttt{err\_corr\_synt}: error on the \textsc{smarty} flux corrected using the synthetic star (erg\,cm$^{-2}$\,s$^{-1}$\,\AA$^{-1}$);
    \item \texttt{flux\_cont\_synt}: the continuum of the synthetic star (erg\,cm$^{-2}$\,s$^{-1}$\,\AA$^{-1}$);
    \item \texttt{flux\_orig}: flux of the original \textsc{smarty} spectrum (erg\,cm$^{-2}$\,s$^{-1}$\,\AA$^{-1}$);
    \item \texttt{err\_orig}: error on the flux of the original \textsc{smarty} spectrum;
    \item \texttt{flux\_cont\_orig}: continuum of the original  \textsc{smarty} spectrum;
    \item \texttt{flux\_corr\_ref}: rest-frame flux of the \textsc{smarty} spectrum corrected using the continuum of the star in common with other libraries \textit{when available} (erg\,cm$^{-2}$\,s$^{-1}$\,\AA$^{-1}$);
    \item \texttt{err\_corr\_ref}: error on the \textsc{smarty} spectrum corrected using the star in common with other libraries \textit{when available};
    \item \texttt{flux\_cont\_ref}: continuum of the star in common with other libraries \textit{when available} (erg\,cm$^{-2}$\,s$^{-1}$\,\AA$^{-1}$).
\end{enumerate}



\bibliographystyle{mnras}
\bibliography{references} 




\appendix


\section{Continuum fit}
\label{Ap:cont_fit}

\begin{figure*}
    \centering
    \begin{tabular}{ccc}
    \vspace{-0.16cm}
    \includegraphics[width=0.99\hsize,page=1]{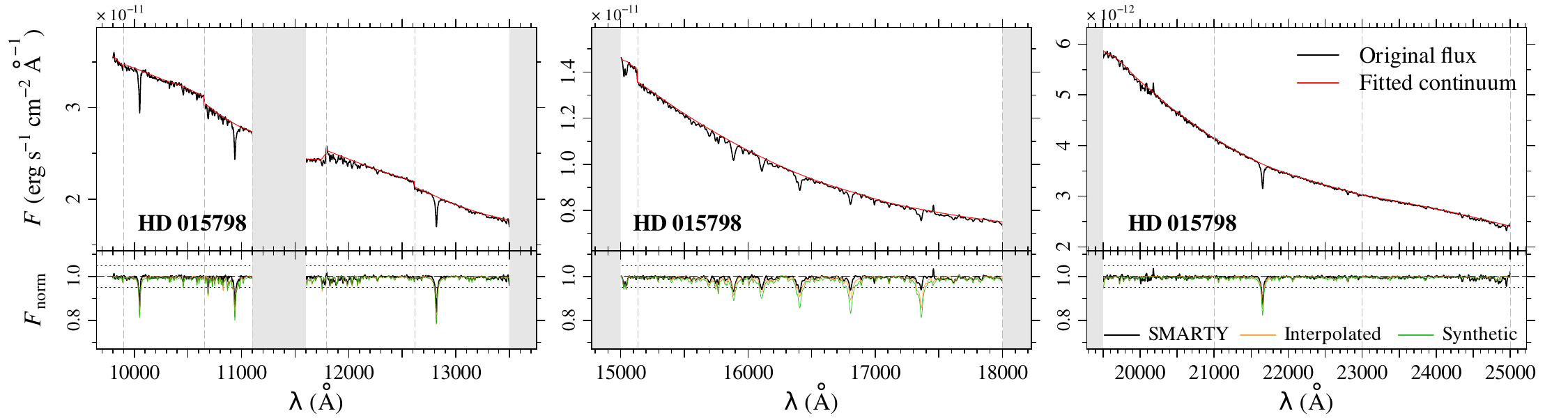}  \\
    \vspace{-0.16cm}
    \includegraphics[width=0.99\hsize,page=1]{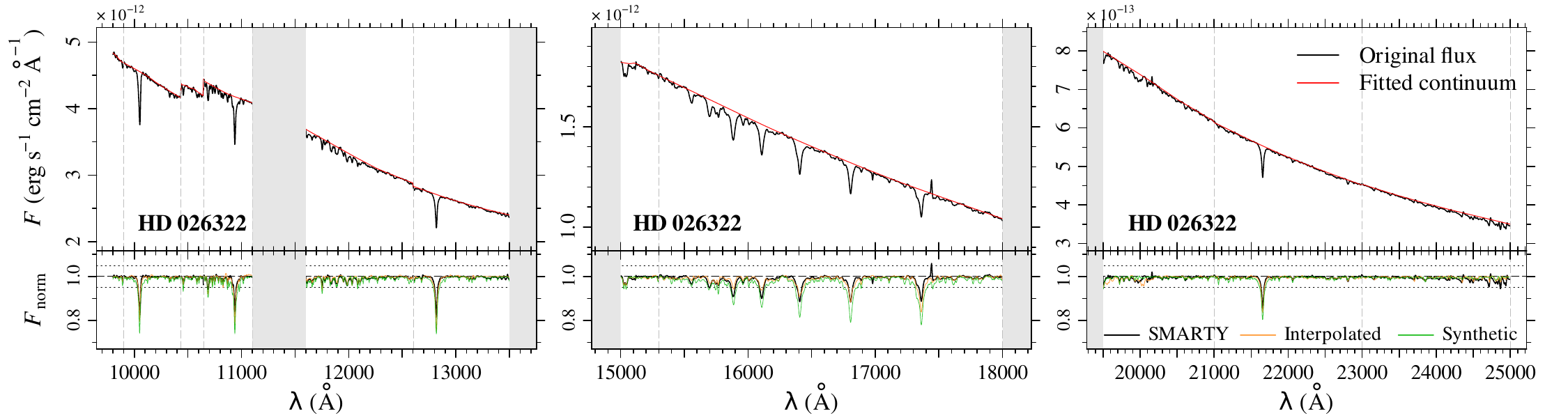}  \\
    \vspace{-0.16cm}
    \includegraphics[width=0.99\hsize,page=1]{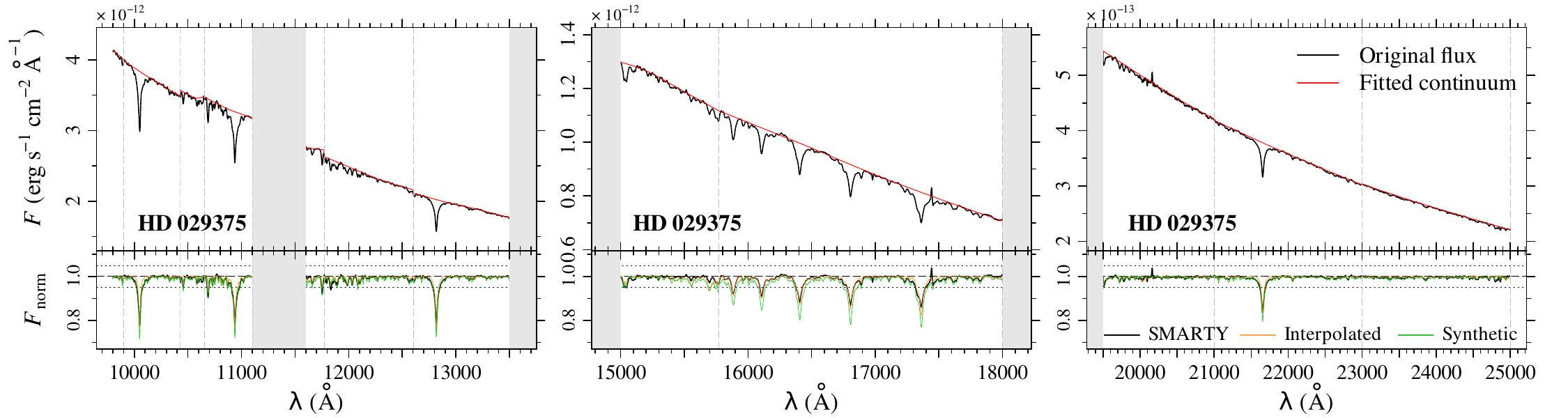}  \\
    \vspace{-0.16cm}
    \includegraphics[width=0.99\hsize,page=1]{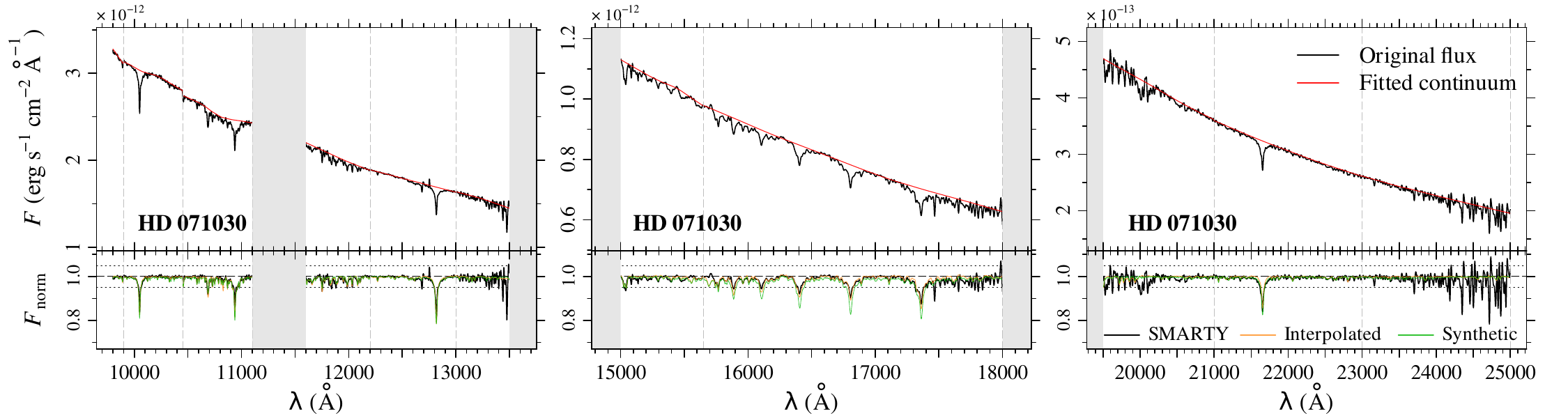}  \\
    \includegraphics[width=0.99\hsize,page=1]{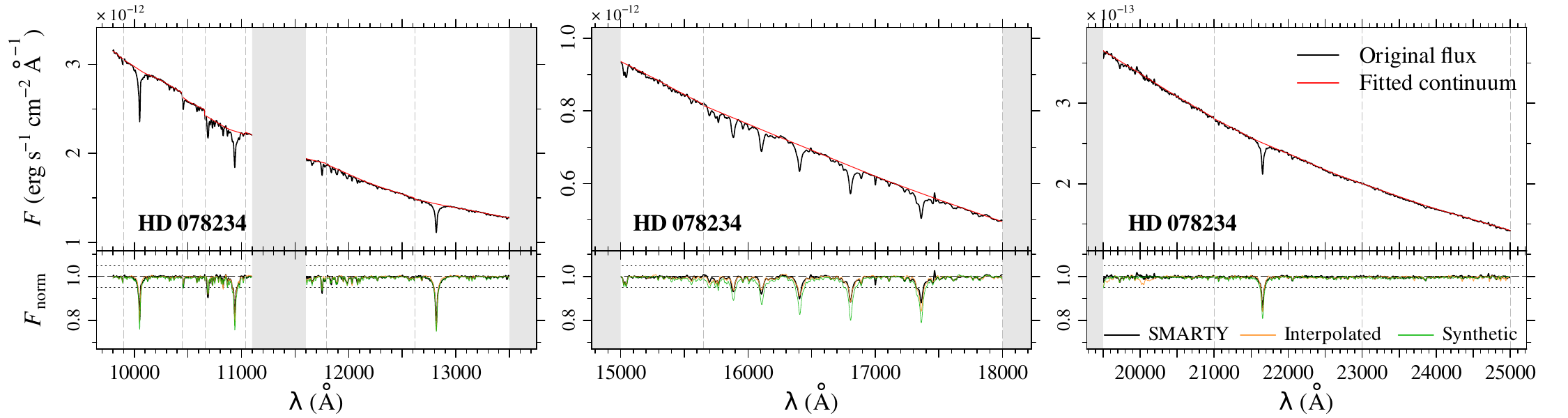}  \\
    \end{tabular}
    \caption{Illustration of the continuum-fit process. The notation is the same as in Fig.~\ref{fig:cont_fit}.}
    \label{fig:cont_fit1}
\end{figure*}

\begin{figure*}
    \centering
    \ContinuedFloat
    \begin{tabular}{ccc}
    \vspace{-0.16cm}
    \includegraphics[width=0.99\hsize,page=1]{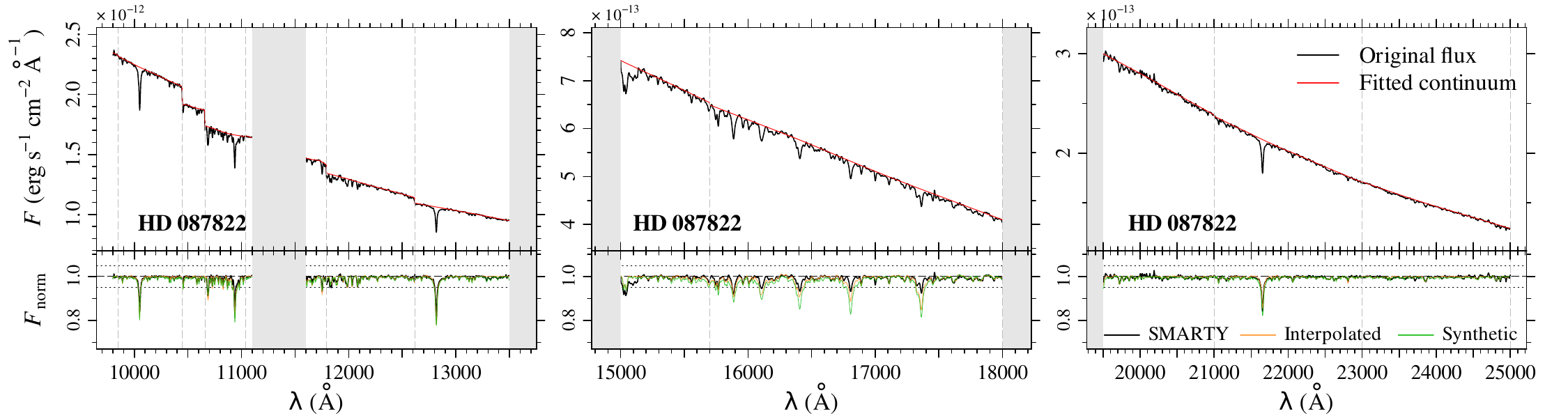}  \\
    \vspace{-0.16cm}
    \includegraphics[width=0.99\hsize,page=1]{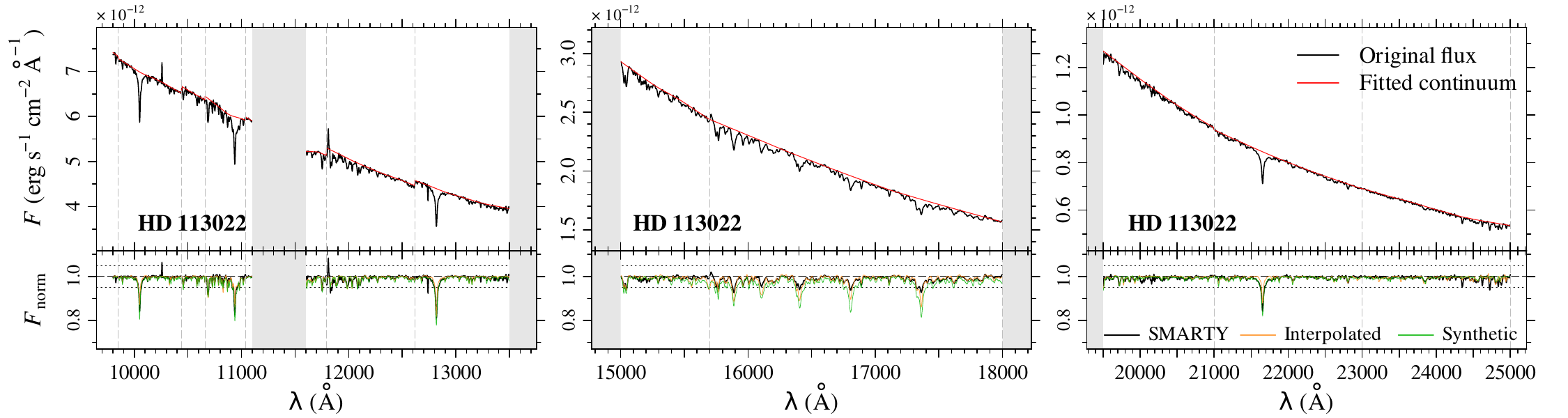}  \\
    \vspace{-0.16cm}
    \includegraphics[width=0.99\hsize,page=1]{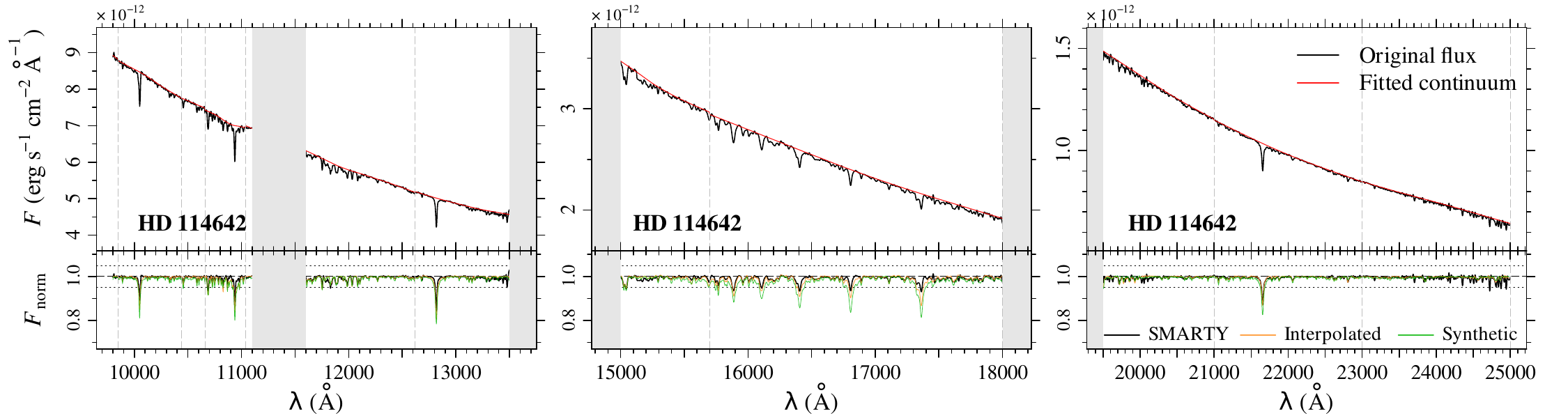} \\
    \vspace{-0.16cm}
    \includegraphics[width=0.99\hsize,page=1]{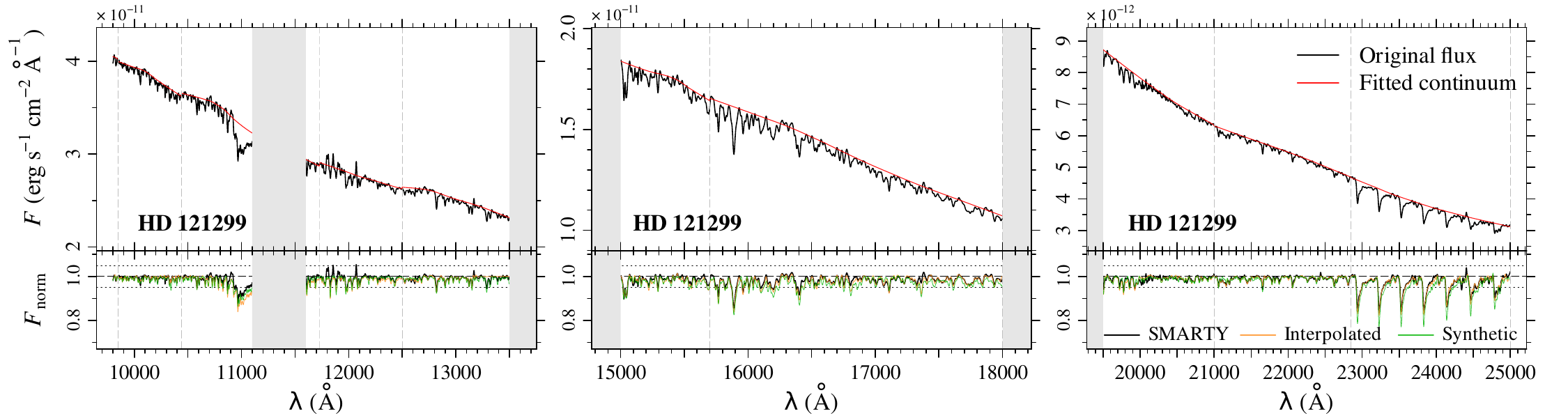} \\
    \includegraphics[width=0.99\hsize,page=1]{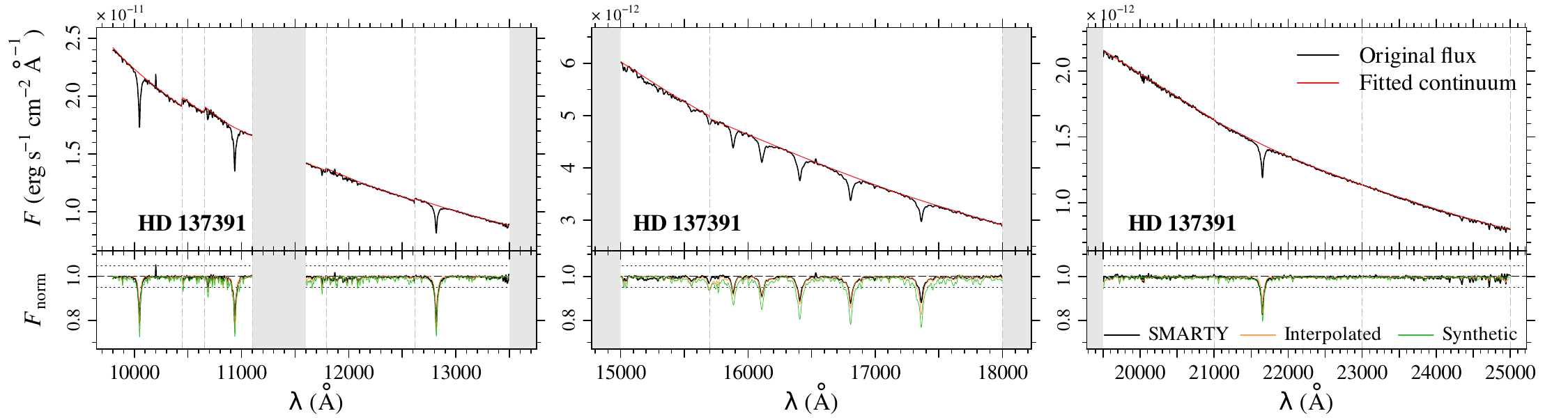} \\
    \end{tabular}
    \caption{Illustration of the continuum-fit process. The notation is the same as in Fig.~\ref{fig:cont_fit}.}
\end{figure*}

\begin{figure*}
    \centering
    \ContinuedFloat
    \begin{tabular}{ccc}
    \vspace{-0.16cm}
    \includegraphics[width=0.99\hsize,page=1]{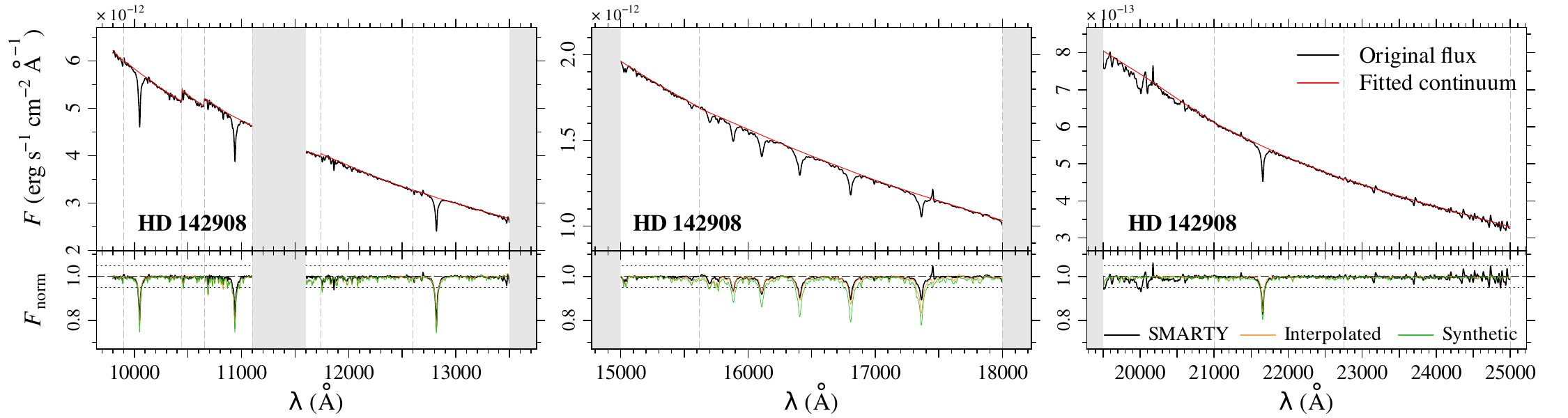} \\
    \vspace{-0.16cm}
    \includegraphics[width=0.99\hsize,page=1]{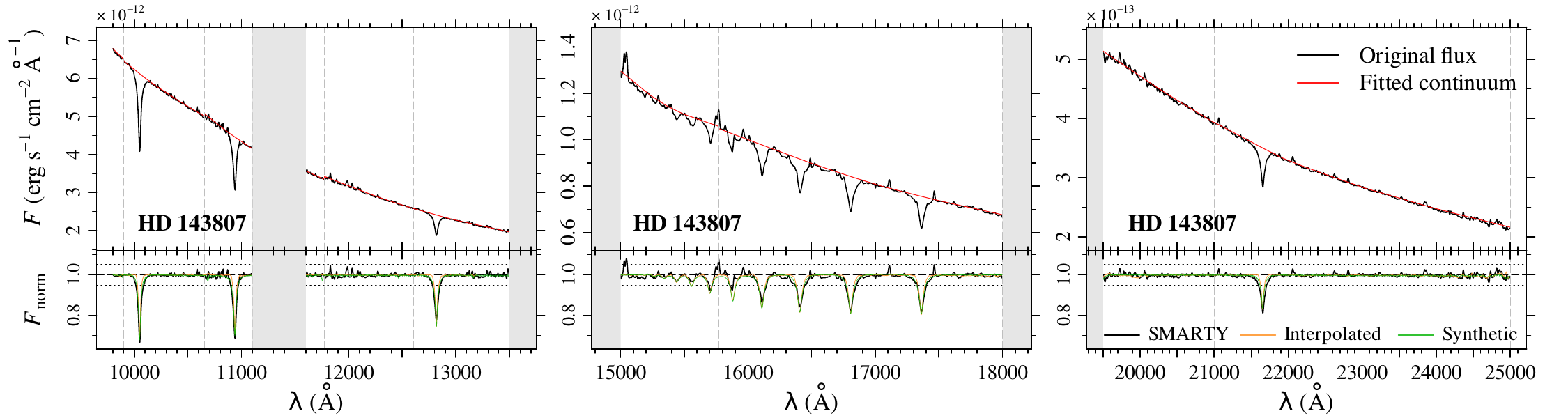} \\
    \vspace{-0.16cm}
    \includegraphics[width=0.99\hsize,page=1]{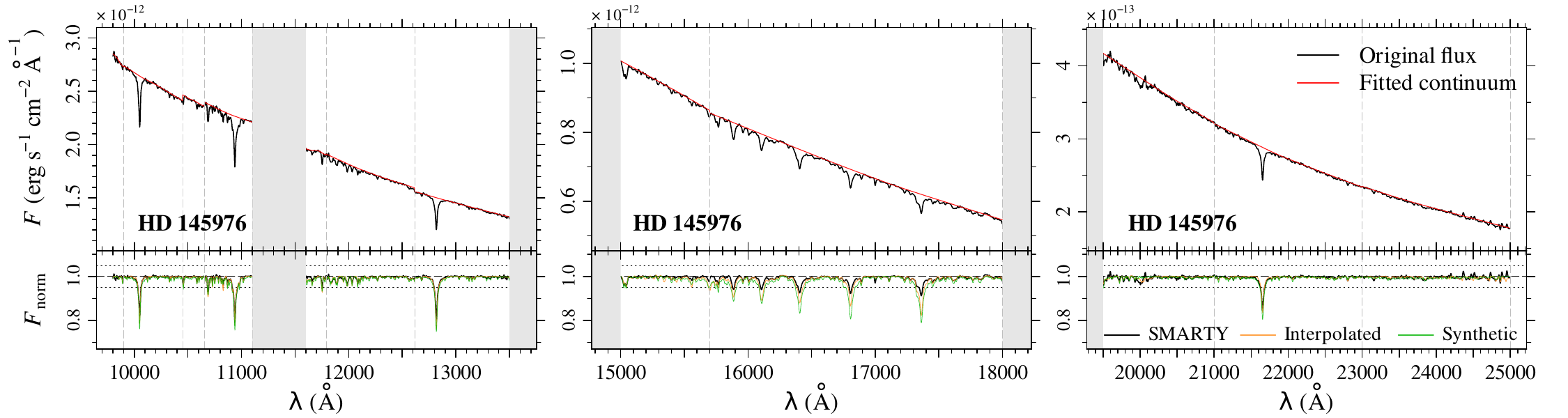} \\
    \vspace{-0.16cm}
    \includegraphics[width=0.99\hsize,page=1]{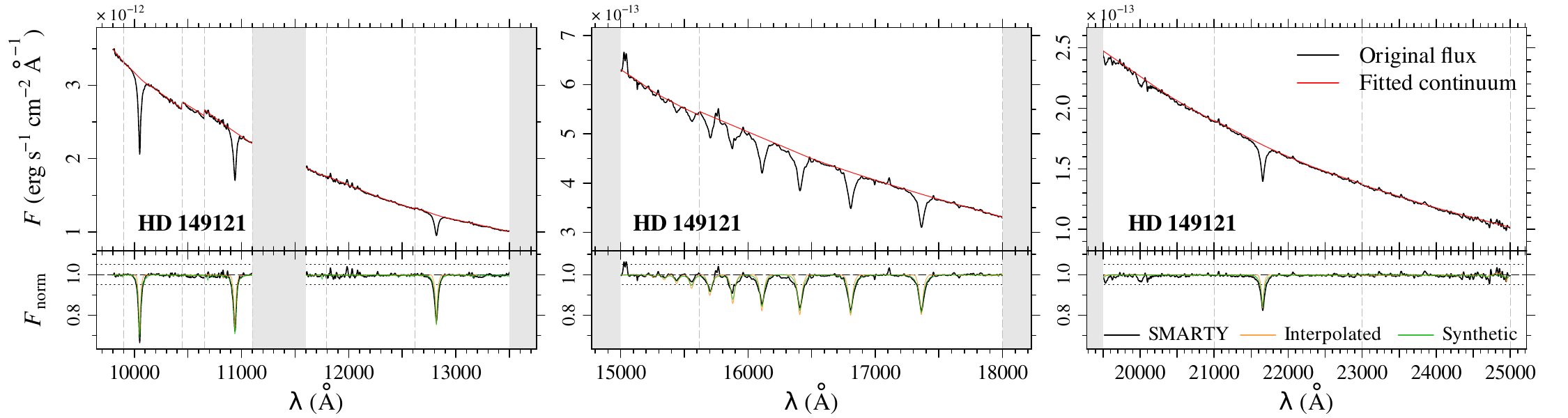} \\
    \includegraphics[width=0.99\hsize,page=1]{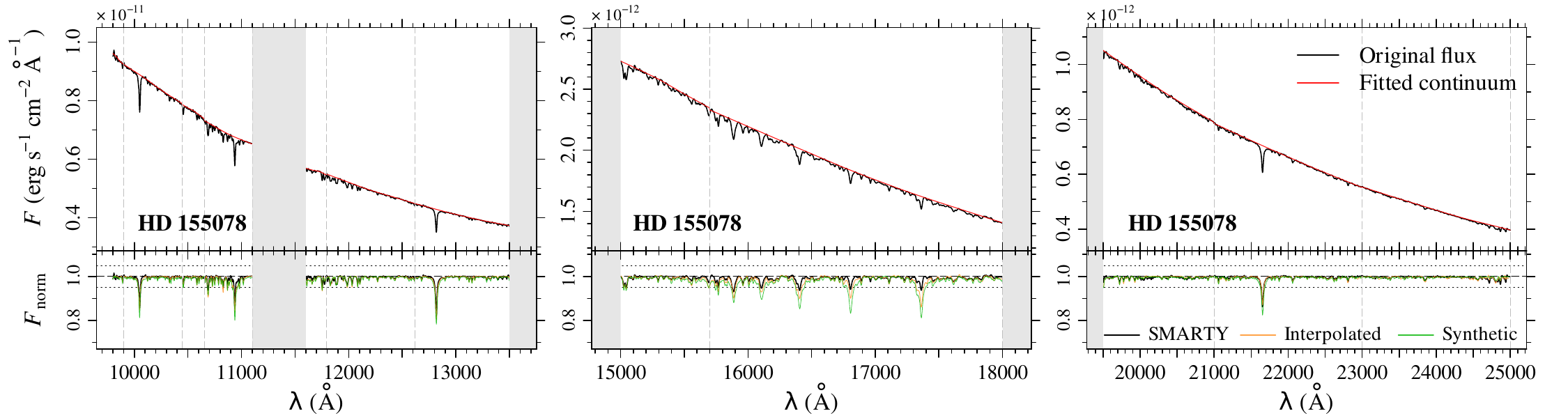} \\
    \end{tabular}
    \caption{Illustration of the continuum-fit process. The notation is the same as in Fig.~\ref{fig:cont_fit}.}
\end{figure*}

\begin{figure*}
    \centering
    \ContinuedFloat
    \begin{tabular}{ccc}
    \vspace{-0.16cm}
    \includegraphics[width=0.99\hsize,page=1]{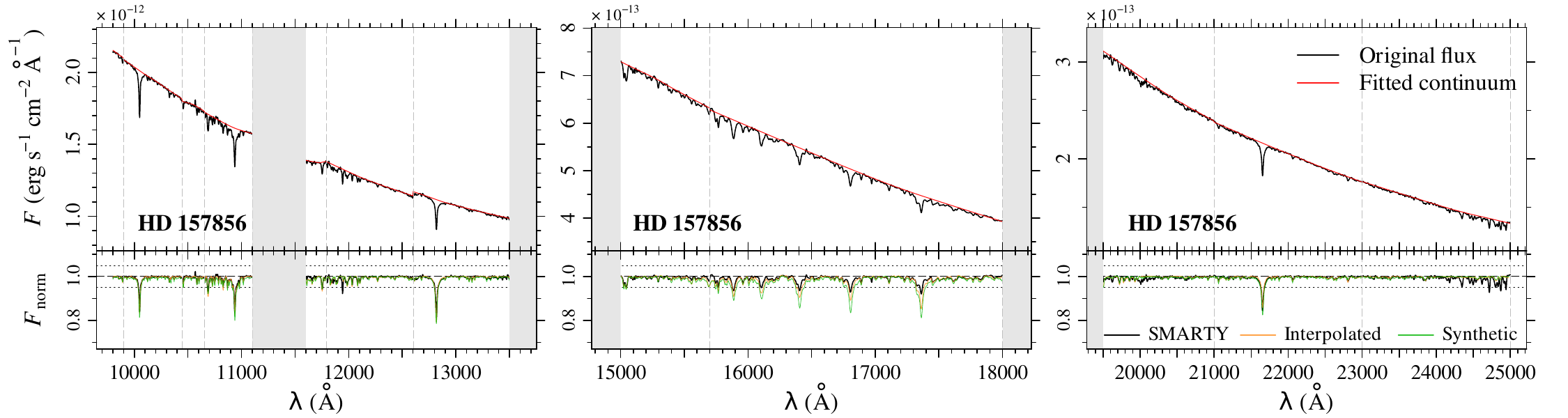} \\
    \vspace{-0.16cm}
    \includegraphics[width=0.99\hsize,page=1]{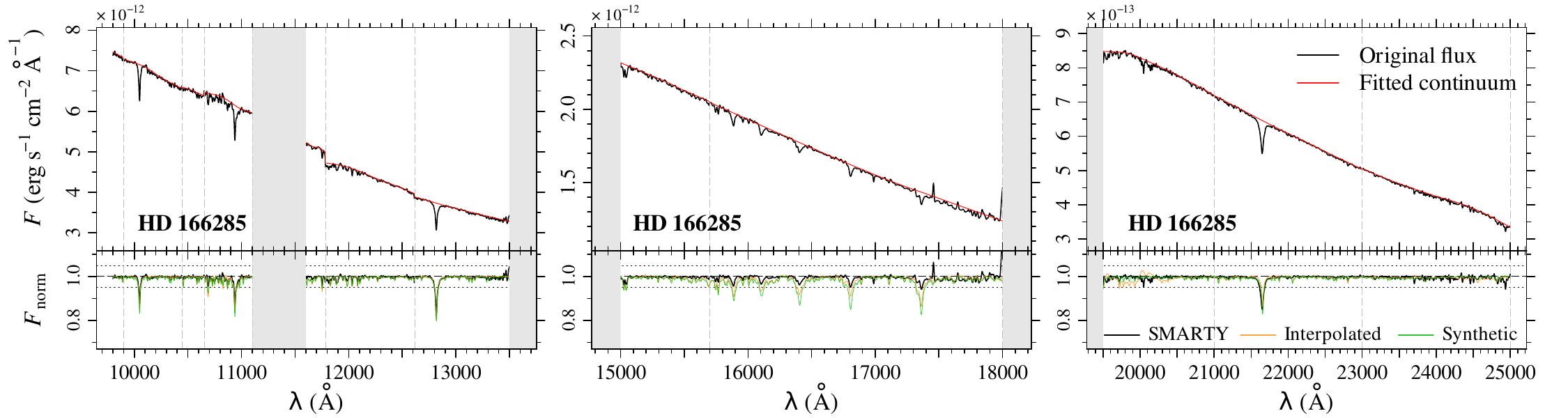} \\
    \vspace{-0.16cm}
    \includegraphics[width=0.99\hsize,page=1]{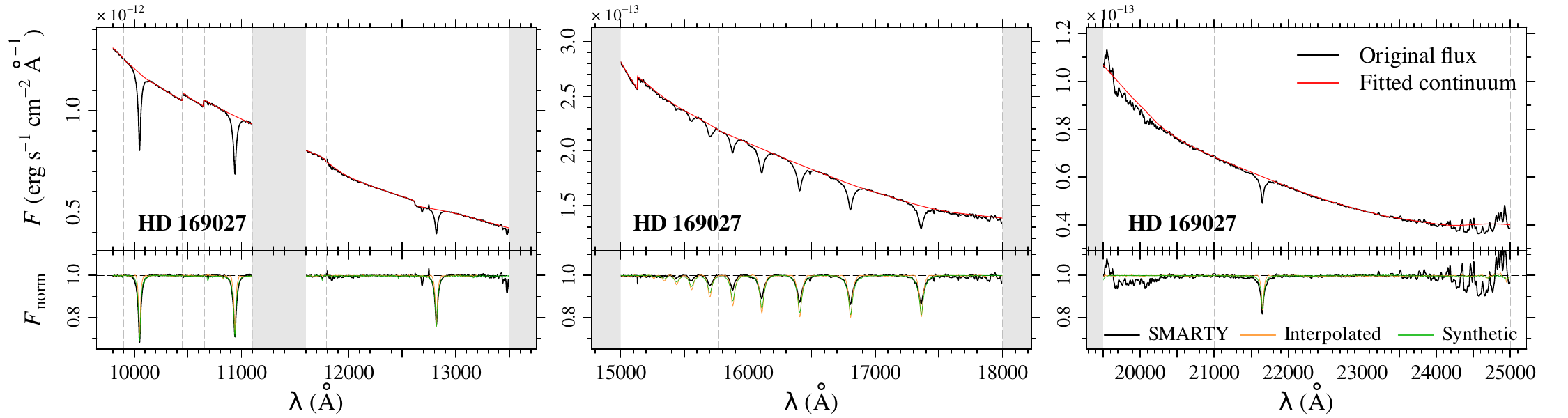} \\
    \vspace{-0.16cm}
    \includegraphics[width=0.99\hsize,page=1]{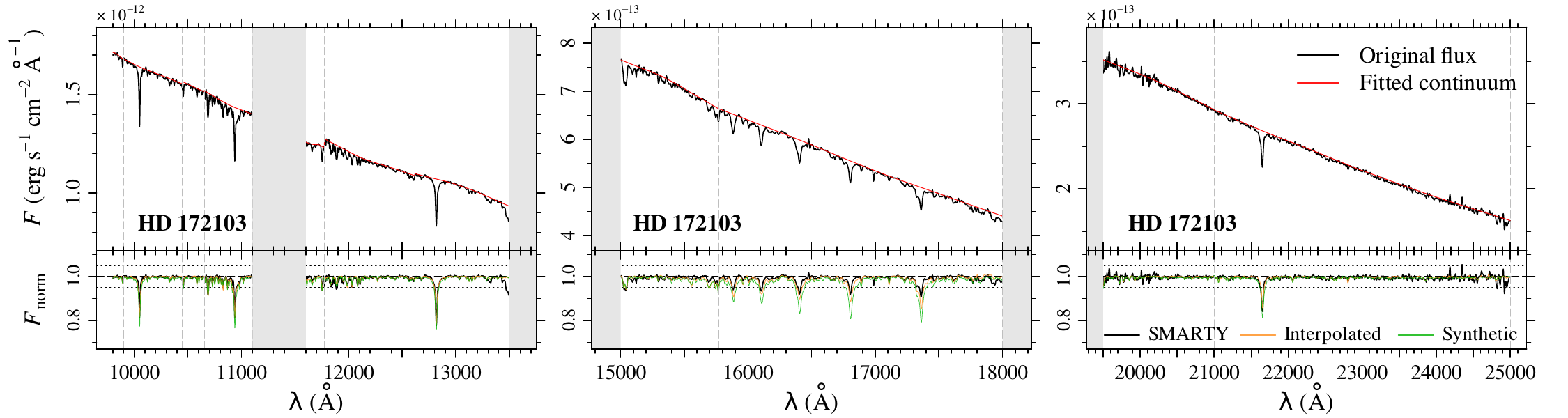} \\
    \includegraphics[width=0.99\hsize,page=1]{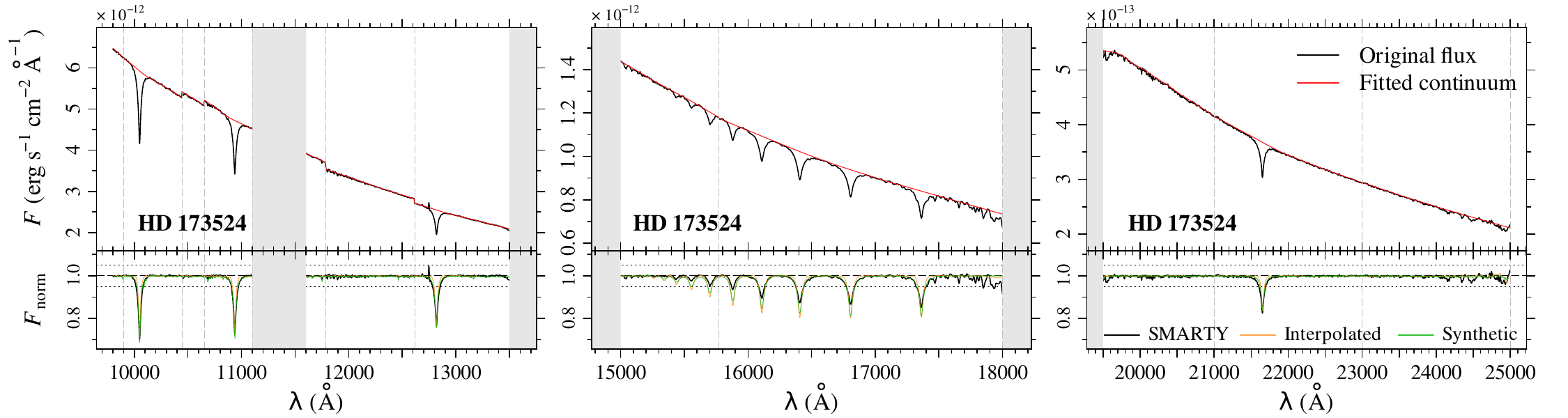} \\
    \end{tabular}    
    \caption{Illustration of the continuum-fit process. The notation is the same as in Fig.~\ref{fig:cont_fit}.}
\end{figure*}

\begin{figure*}
    \centering
    \ContinuedFloat
    \begin{tabular}{ccc}
    \vspace{-0.16cm}
    \includegraphics[width=0.99\hsize,page=1]{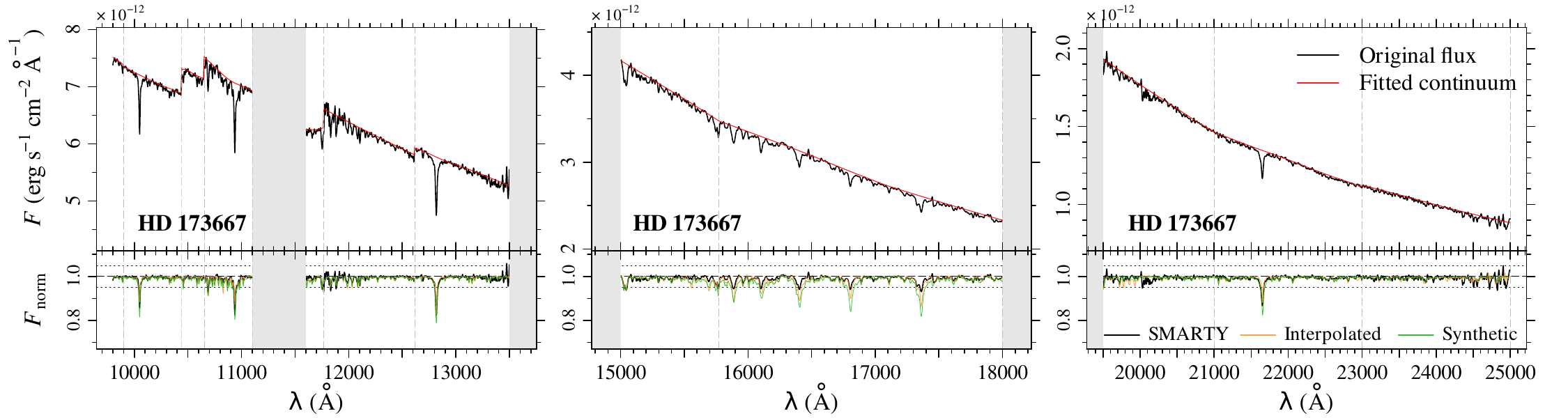} \\
    \vspace{-0.16cm}
    \includegraphics[width=0.99\hsize,page=1]{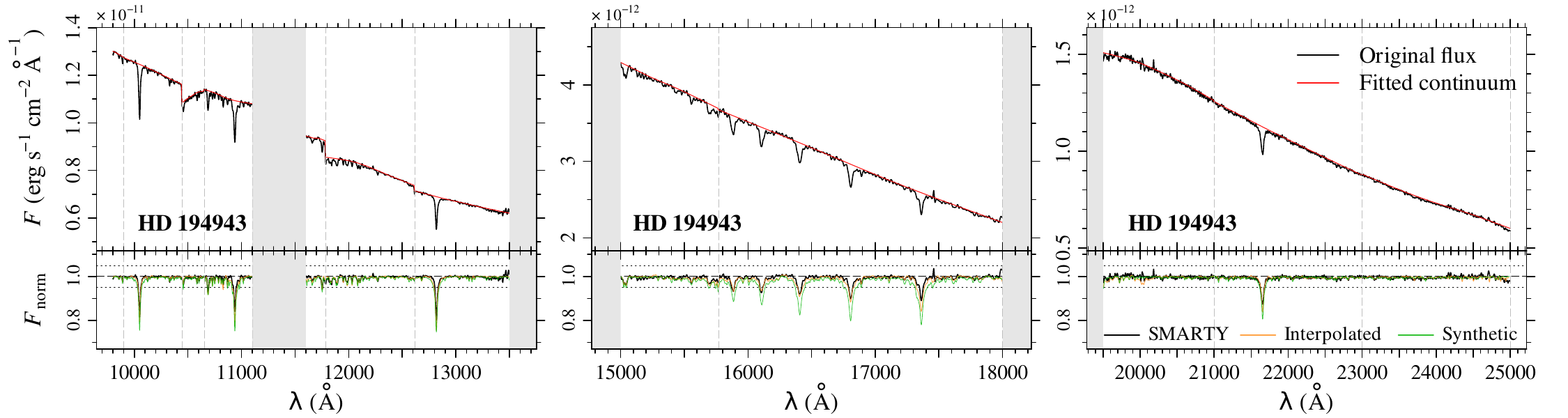} \\
    \vspace{-0.16cm}
    \includegraphics[width=0.99\hsize,page=1]{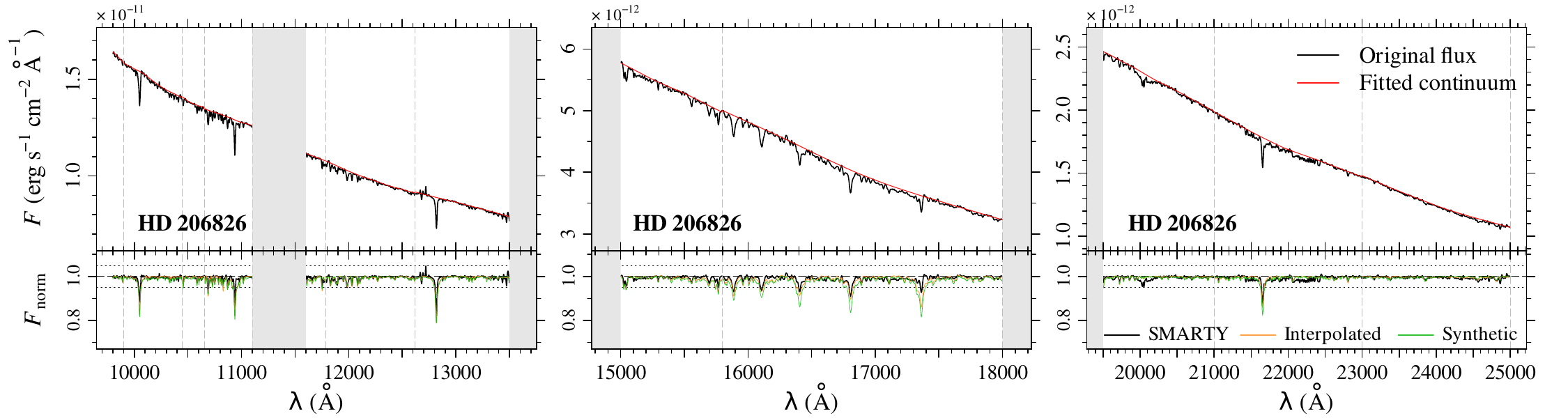} \\
    \vspace{-0.16cm}
    \includegraphics[width=0.99\hsize,page=1]{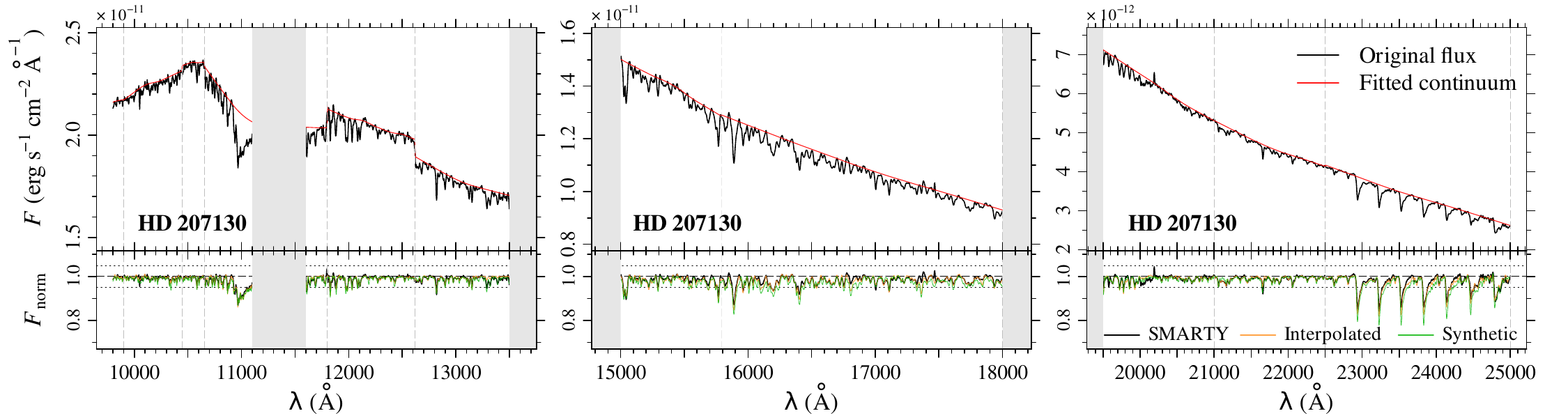} \\
    \includegraphics[width=0.99\hsize,page=1]{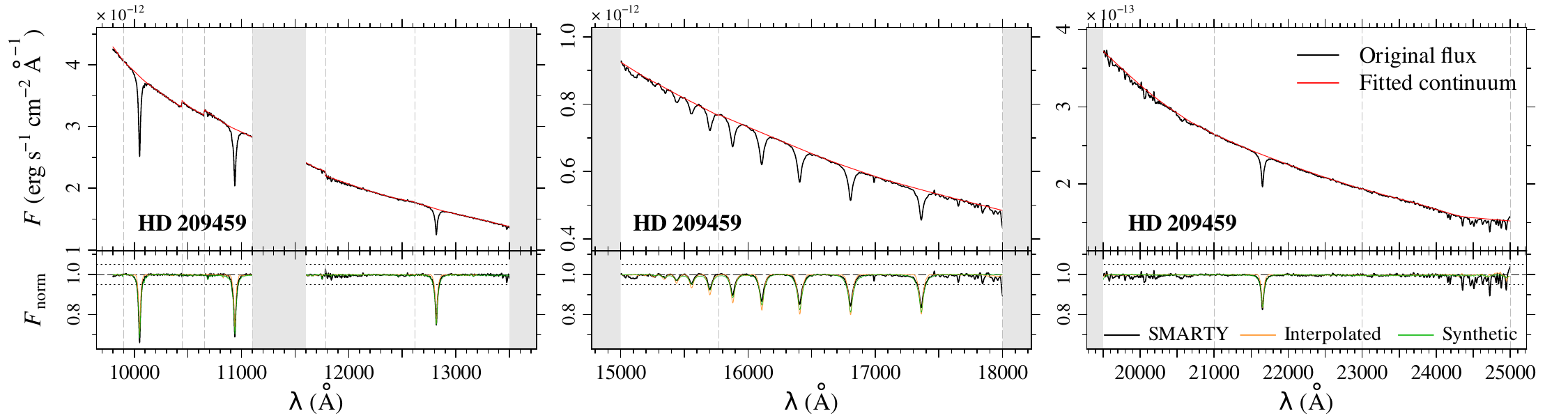} \\
    \end{tabular}
    \caption{Illustration of the continuum-fit process. The notation is the same as in Fig.~\ref{fig:cont_fit}.}
\end{figure*}

\begin{figure*}
    \centering
    \ContinuedFloat
    \begin{tabular}{ccc}
    \vspace{-0.16cm}
    \includegraphics[width=0.99\hsize,page=1]{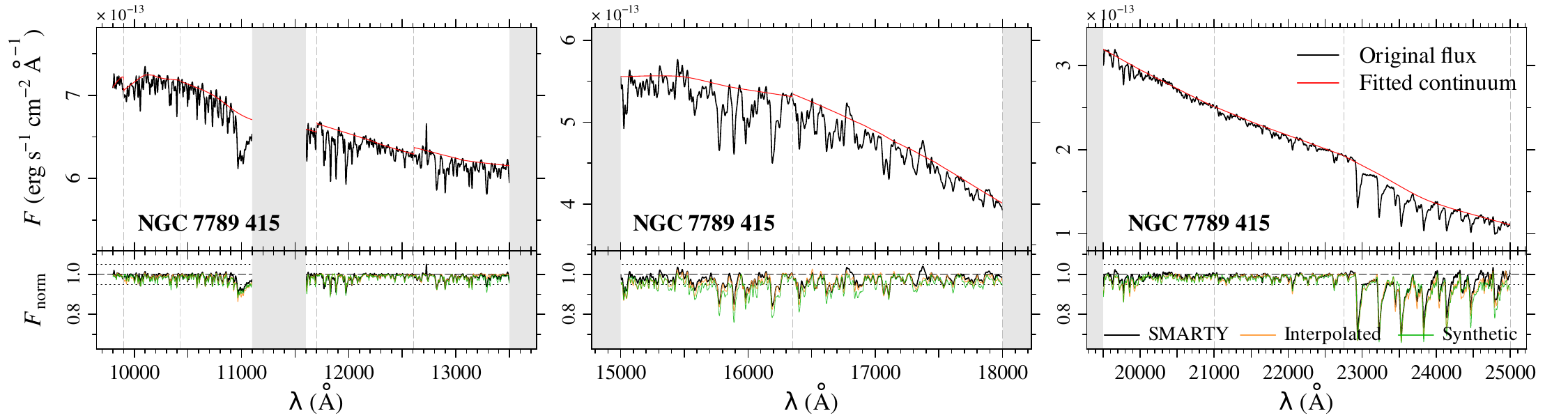} \\
    \vspace{-0.16cm}
    \includegraphics[width=0.99\hsize,page=1]{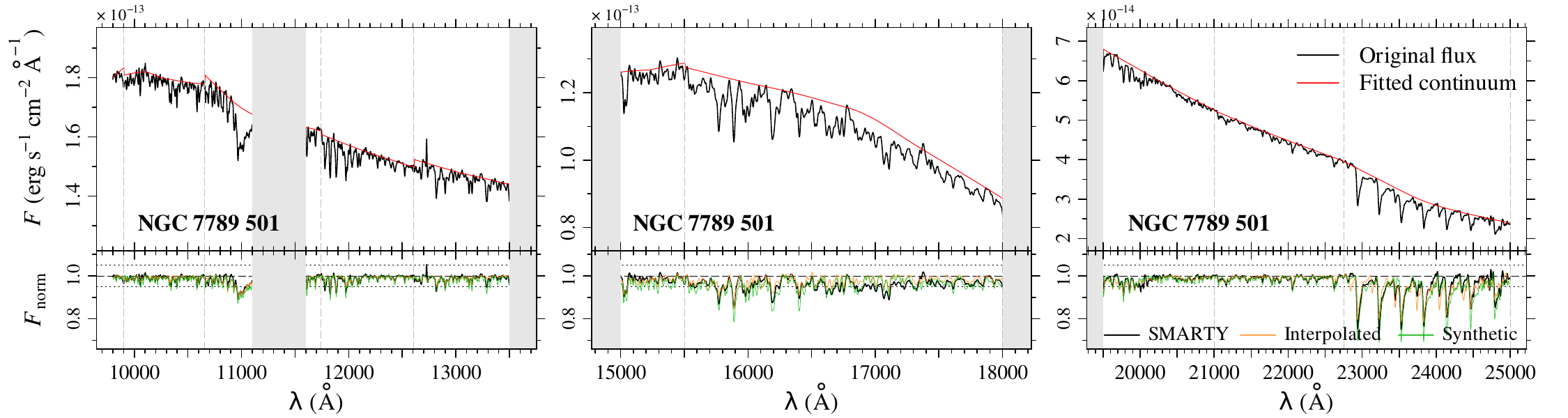} \\
    \vspace{-0.16cm}
    \includegraphics[width=0.99\hsize,page=1]{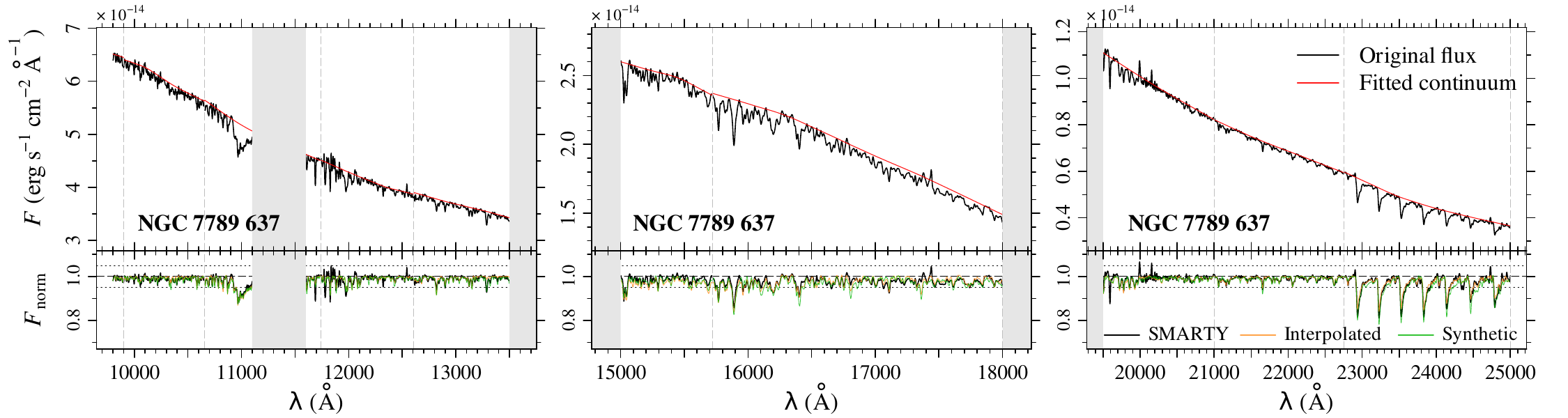} \\
    \includegraphics[width=0.99\hsize,page=1]{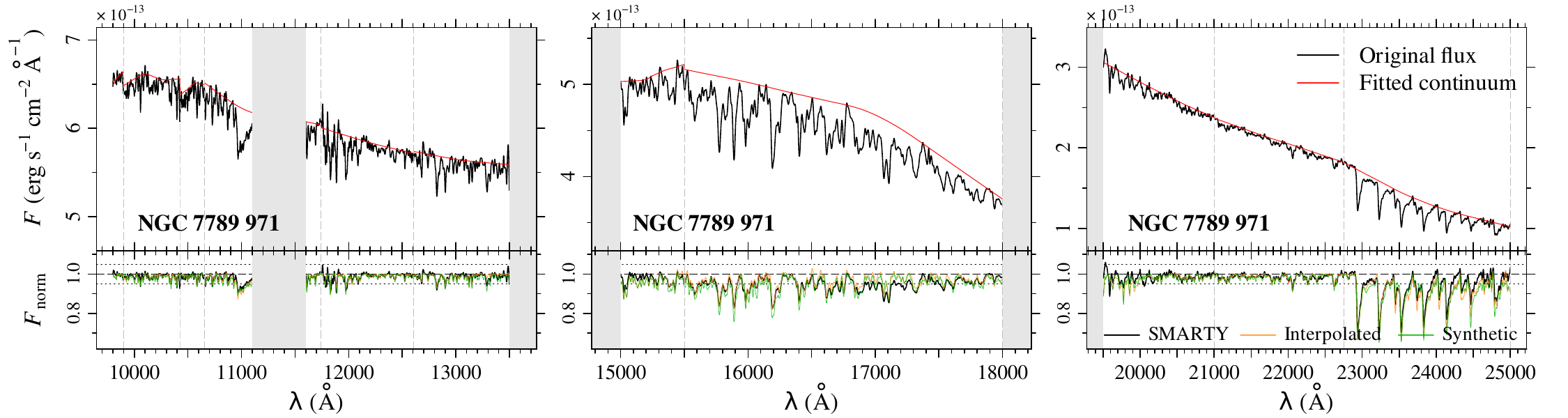} \\
    \end{tabular}
    \caption{Illustration of the continuum-fit process. The notation is the same as in Fig.~\ref{fig:cont_fit}.}
\end{figure*}


\bsp	
\label{lastpage}
\end{document}